\documentclass{jetp} 

\usepackage{graphicx}
\usepackage{epsfig}
\usepackage[T1,T2A]{fontenc}
\usepackage[cp1251]{inputenc}
\usepackage[english]{babel}
\usepackage[russian]{babel}
\usepackage{amssymb}
\usepackage{amsmath}
\usepackage{setspace}
\usepackage{cuted}
 \usepackage{epigraph}

\def\beq{\begin{equation}}
\def\eeq{\end{equation}}
\def\Li{LiCu$_2$O$_2$}
\def\LiV{LiCuVO$_4$}
\def\Cs{CsCuCl$_3$}
\def\Y{YFeO$_3$}

\begin{document}
\English

\title{Dzyaloshinskii interaction and  exchange-relativistic effects in orthoferrites
} 




\author{A.~S.}{Moskvin} 
\email{alexander.moskvin@urfu.ru} 
\affiliation{Ural Federal University, 620083, Ekaterinburg, Russia
} 
\affiliation{Institute of Metal Physics UB RAS, 620108, Ekaterinburg, Russia} 







\abstract{We present an  overview of the microscopic theory of the Dzyaloshinskii-Moriya (DM) coupling and related exchange-relativistic effects such as exchange anisotropy, electron-nuclear antisymmetric supertransferred hyperfine  interactions, antisymmetric magnetogyrotropic effects,
and antisymmetric magnetoelectric coupling  in strongly correlated 3$d$ compounds focusing on orthoferrites RFeO$_3$ (R is a rare-earth ion or yttrium Y).
Most attention in the paper centers around the derivation of the Dzyaloshinskii vector, its value, orientation, and sense (sign)  under different types of the (super)exchange interaction and crystal field. Microscopically derived expression for the dependence of the Dzyaloshinskii vector on the superexchange geometry allows one to find all the overt and hidden canting angles in orthoferrites RFeO$_3$ as well as corresponding contribution to magnetic anisotropy. Being based on the theoretical predictions regarding the sign of the Dzyaloshinskii vector we have predicted and study in detail a novel magnetic phenomenon, {\it weak ferrimagnetism} in mixed weak ferromagnets with competing signs of the Dzyaloshinskii vectors. The ligand NMR  measurements in weak ferromagnets are shown to be an effective tool to inspect the effects of DM coupling in an external magnetic field.
Along with orthoferrites RFeO$_3$ and weak ferrimagnets RFe$_{1-x}$Cr$_x$O$_3$, although to a lesser extent, we address such typical weak ferromagnets as
$\alpha$-Fe$_2$O$_3$, FeBO$_3$, and FeF$_3$.
}

\maketitle


\section{Introduction}




It is not often in the history of science that one paper of an author opens up a novel  field of theoretical and experimental research. This is exactly what happened with the article by Igor Dzyaloshinskii\,\cite{Dzyaloshinskii}, devoted to the explanation of the phenomenon of weak ferromagnetism.

More than a hundred years have passed after T. Smith\,\cite{Smith} found in 1916 a weak, or $parasitic$ ferromagnetism in an "international family line" of different natural hematite $\alpha$-Fe$_2$O$_3$ single crystalline samples from Italy, Hungary, Brasil, and Russia (Schabry, Ural Mountains, near Ekaterinburg) that was assigned to ferromagnetic impurities. Later the phenomenon was observed in many other materials, in fluoride NiF$_2$ with rutile structure, in the orthorhombic orthoferrites RFeO$_3$ (where R is a rare-earth element or Y), in rhombohedral antiferromagnets MnCO$_3$, NiCO$_3$, CoCO$_3$, and FeBO$_3$. However, only in 1954 L.~M. Matarrese and J.~W. Stout for NiF$_2$\,\cite{Matarrese} and in 1956 A.~S. Borovik-Romanov and M.~P. Orlova for very pure synthesised  carbonates  MnCO$_3$ and CoCO$_3$\,\cite{Borovik} have firmly established that  weak ferromagnetism is observed in chemically pure antiferromagnetic materials and therefore it is an intrinsic property of some antiferromagnets, the connexion between the weak ferromagnetism and any impurities or inhomogeneities seems very unlikely. Furthermore, Borovik-Romanov and Orlova assigned the uncompensated moment in MnCO$_3$ and CoCO$_3$ to an overt canting of the two magnetic sublattices in almost antiferromagnetic matrix.
 The model of a canted  antiferromagnet   became generally adopted model of the weak ferromagnet.



A theoretical explanation and first thermodynamic theory for weak ferromagnetism in $\alpha$-Fe$_2$O$_3$, MnCO$_3$, and CoCO$_3$ was provided by I.~E. Dzyaloshinskii (Dzialoshinskii, Dzyaloshinsky)\,\cite{Dzyaloshinskii} in 1957 on the basis of symmetry considerations and Landau's theory of phase transitions of the second kind.

\begin{figure}[t]
\centering
\includegraphics[width=8.5cm,angle=0]{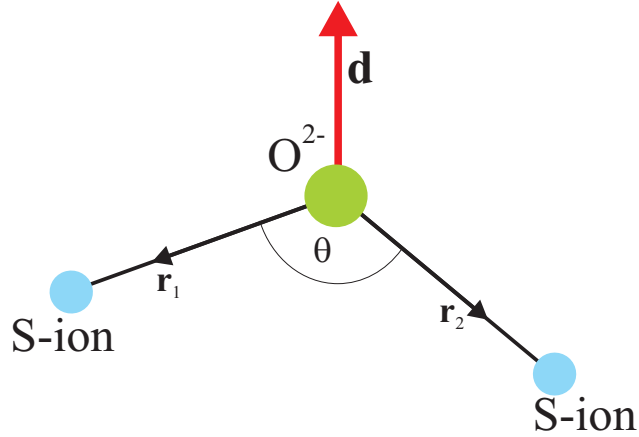}
\caption{(Color online) Superexchange geometry and the Dzyaloshinskii vector.}
\label{fig1}
\end{figure}

 Free energy of the two-sublattice uniaxial weak ferromagnet such as $\alpha$-Fe$_2$O$_3$, MnCO$_3$, CoCO$_3$, FeBO$_3$ was shown to be written as follows
\[
F=MH_E({\bf m}_1\cdot {\bf m}_2)-M{\bf H}_0({\bf m}_1+{\bf m}_2)+E_D+E_A
\]
\begin{equation}
	= MH_E({\bf m}^2-{\bf l}^2)-M{\bf H}_0{\bf m}+E_D+E_A \, .
\end{equation}
In this expression ${\bf m}_1$ and ${\bf m}_2$ are unit vectors in the directions of the sublattice moments, $M$ is the sublattice
magnetization, ${\bf m}=\frac{1}{2}({\bf m}_1+{\bf m}_2)$ and  ${\bf l}=\frac{1}{2}({\bf m}_1-{\bf m}_2)$ are the ferro- and antiferromagnetic vectors, respectively, $H_0$ is the applied field, $H_E$ is the exchange field,
$$
	E_D=-MH_D[{\bf m}_1\times {\bf m}_2]_z=+2MH_D[{\bf m}\times {\bf l}]_z= \\
$$
\begin{equation}
+2MH_D(m_xl_y-m_yl_x)
\end{equation}
is now called the Dzyaloshinskii interaction, $H_D$\,$>$\,0 is the Dzyaloshinskii field.
The anisotropy energy $E_A$ is assumed to have the form: $E_A = H_A/2M(m_{1z}^2+m_{2z}^2)=2H_A/2M(m_{z}^2+l_{z}^2)$, where $H_A$ is the anisotropy field. The choice of sign for the anisotropy field $H_A$ assumes that the $c$ axis is a hard direction of magnetization.
In a general sense the Dzyaloshinskii interaction implies the terms that are linear both on  ferro- and antiferromagnetic vectors. For instance, in orthorhombic orthoferrites and orthochromites the Dzyaloshinskii interaction consists of the antisymmetric and symmetric terms
$$
	E_D= d_1m_zl_x+d_2m_xl_z=
$$
$$
\frac{d_1-d_2}{2}(m_zl_x-m_xl_z)+\frac{d_1+d_2}{2}(m_zl_x+m_xl_z)=
$$
\begin{equation}
	-2MH_D[{\bf m}\times {\bf l}]_y+\frac{d_1+d_2}{2}(m_zl_x+m_xl_z)\, ,
\end{equation}
while for tetragonal fluorides NiF$_2$ and CoF$_2$ the Dzyaloshinskii interaction consists of the only symmetric term.
Despite Dzyaloshinskii supposed that weak ferromagnetism is due to relativistic spin-lattice and magnetic dipole interaction,  the theory was phenomenological one and did not clarify the microscopic nature of the Dzyaloshinskii interaction that does result in the canting.
 Later on, in 1960, T. Moriya\,\cite{Moriya} suggested a model microscopic theory of the exchange-relativistic antisymmetric exchange interaction to be a main contributing mechanism of weak ferromagnetism

\begin{equation}
V_{DM}=\sum_{mn}({\bf d}_{mn}\cdot\left[{\bf S}_m\times{\bf S}_n\right]) \, ,
	\label{DM}
\end{equation}
now called Dzyaloshinskii-Moriya (DM) spin coupling. Here,  ${\bf d}_{mn}$ is the axial
 Dzyaloshinskii vector.
Presently Keffer\,\cite{Keffer} proposed a simple phenomenological expression for the Dzyaloshinskii vector for two magnetic ions M$_i$ and M$_j$ interacting by the superexchange mechanism via intermediate ligand O (see Fig.\,\ref{fig1}):
\begin{equation}
	\mathbf{d}_{ij} \propto [\mathbf{r}_i \times \mathbf{r}_j] \, ,
\end{equation}
where ${\bf r}_{i,j}$ are unit radius vectors for O\,-\,M$_{i,j}$ bonds with presumably equal bond lenghts. Later on Moskvin\,\cite{1971} derived a microscopic formula for Dzyaloshinskii vector
 \begin{equation}
	\mathbf{d}_{ij} = d_{ij}(\theta) [\mathbf{r}_i \times \mathbf{r}_j] \, ,
	\label{d12}
\end{equation}
where
\begin{equation}
	d_{ij}(\theta)=d_1(R_i,R_j)+d_2(R_i,R_j)cos\theta_{ij}
\end{equation}
with $\theta_{ij}$ being the M$_{i}$\,-\,O\,-\,M$_{j}$ bonding angle.
The sign of the scalar parameter $d_{ij}(\theta)$ can be addressed to be the sign (sense) of the Dzyaloshinskii vector.  The formula (\ref{d12}) was shown to work only for $S$-type magnetic ions with orbitally nondegenerate ground state, e.g. for 3$d$-ions with half-filled shells (3$d^5$ or $t_{2g}^3$, $t_{2g}^3e_g^2$, $t_{2g}^6e_g^2$).

It should be noted that sometimes instead of (\ref{d12}) one may use another form of the structural factor for the Dzyaloshinskii vector:
\begin{equation}
	\left[{\bf r}_1\times{\bf r}_2\right]=\frac{1}{2}\left[({\bf r}_1-{\bf r}_2)\times({\bf r}_1+{\bf r}_2)\right]=\frac{1}{2l^2}\left[{\bf R}_{12}\times{\pmb \rho}_{12
}\right] \, ,
\label{Rrho}
\end{equation}
where ${\bf R}_{12}={\bf R}_1-{\bf R}_2$, ${\pmb \rho}_{12}=({\bf R}_1+{\bf R}_2)$.

\begin{table*}
\caption{Main exchange and DM coupling parameters in orthoferrites compared with other weak ferromagnets (WFMs),
$I$ is the exchange integral, $\alpha_D$ is the canting angle. See text for detail.}
\begin{tabular}{|c|c|c|c|c|c|c|c|c|}
\hline
 WFM & R$_{FeO}$, \AA & $\theta$ & T$_N$, K & $I$, K (MFA) & H$_E$, Tesla & $\alpha_D$ & H$_D$, Tesla & $d$($\theta$), K \\ \hline
YFeO$_3$ & 2.001 (x2) & 145$^{\circ}$ &  640 & 36.6 & 640 & $1.1\cdot 10^{-2}$ & 14 & 3.2\\ \hline
$\alpha$-Fe$_2$O$_3$ & 2.111  & 145$^{\circ}$ &  948 & 54.2 & 870-920 & $1.1\cdot 10^{-3}$ & 1.9-2.2 & 2.3\\ \hline
FeBO$_3$ & 2.028 & 126$^{\circ}$ &  348 & 19.9 & 300 & $1.7\cdot 10^{-2}$ & 10 & 2.3\\ \hline
FeF$_3$ & 1.914  & 153$^{\circ}$ &  363 & 20.7 & 440 & $5.5\cdot 10^{-3}$ & 4.88 & 1.1\\ \hline
 \end{tabular}
\label{D}
\end{table*}

Starting with the pioneering papers by Dzyaloshinskii\,\cite{Dzyaloshinskii} and Moriya\,\cite{Moriya} the DM coupling was extensively investigated in 60-80ths in connection with the weak ferromagnetism focusing on  hematite $\alpha$-Fe$_2$O$_3$ and orthoferrites RFeO$_3$\,\cite{1972,1975,1977,thesis} (see, also review articles Refs.\,\cite{2016,CM-2019}). Typical values of the canting angle $\alpha_D$ turned out to be on the order of 0.001-0.01, in particular, $\alpha_D=1.1\cdot 10^{-3}$ in $\alpha$-Fe$_2$O$_3$\,\cite{Fe2O3}, 2.2-2.9$\cdot 10^{-3}$ in La$_2$CuO$_4$\,\cite{Thio}, $5.5\cdot 10^{-3}$ in FeF$_3$\,\cite{Prozorova},  $1.1\cdot 10^{-2}$ in YFeO$_3$\,\cite{Jacobs}, $1.7\cdot 10^{-2}$ in FeBO$_3$\,\cite{Kotyuzhanskii} (see Table\,\ref{D}).

V.~I. Ozhogin {\it et al}.\,\cite{Ozhogin} in 1968 first  raised the issue of the sign of the Dzyaloshinskii vector, however, only in 1990 the reliable local information on the sign of the Dzyaloshinskii vector, or to be exact, that of the Dzyaloshinskii parameter $d_{12}$, was first extracted from the $^{19}$F ligand NMR (nuclear magnetic resonance) data in weak ferromagnet FeF$_3$\,\cite{sign}.
In 1977 we have shown that the Dzyaloshinskii vectors can be of opposite sign for different pairs of $S$-type ions\,\cite{1977} that allowed us to uncover a novel magnetic phenomenon, {\it weak ferrimagnetism}, and a novel class of magnetic materials, {\it weak ferrimagnets}, which are systems such as solid solutions YFe$_{1-x}$Cr$_x$O$_3$ with competing signs of the Dzyaloshinskii vectors and the very unusual concentration and temperature dependence of the magnetization\cite{WFIM-1}.
The relation between Dzyaloshinskii vector and the superexchange geometry (\ref{d12}) allowed us to find numerically all the overt and hidden canting angles in the rare-earth orthoferrites\,\cite{1975} that was nicely confirmed in $^{57}$Fe NMR\,\cite{Luetgemeier} and neutron diffraction\,\cite{Plakhtii} measurements.

The stimulus to a renewed interest to the subject was given by the cuprate problem, in particular, by the weak ferromagnetism observed in the parent cuprate La$_2$CuO$_4$\,\cite{Thio} and  many other interesting effects for the DM systems, in particular, the "field-induced gap"\, phenomena\,\cite{Affleck}.  At variance with typical 3D systems such as orthoferrites, the cuprates  are characterised by a low-dimensionality, large diversity of Cu-O-Cu bonds including corner- and edge-sharing, different ladder configurations, strong quantum effects for $s=1/2$ Cu$^{2+}$ centers, and a particularly strong Cu-O covalency resulting in a comparable magnitude of hole charge/spin densities on copper and oxygen sites.
Several groups (see, e.g., Refs.\cite{Coffey,Koshibae,Shekhtman}) developed the microscopic model approach by Moriya for different 1D and 2D cuprates, making use of different perturbation schemes, different types of low-symmetry crystalline field, different approaches to intra-atomic electron-electron repulsion. However, despite a rather large number of publications and hot debates (see, e.g., Ref.\cite{debate}) the problem of exchange-relativistic effects, that is of antisymmetric exchange and related problem of spin anisotropy in cuprates remains to be open (see, e.g., Refs.\cite{CM-2019,Tsukada,Kataev,JETP-2007} for recent experimental data and discussion). Common shortcomings of current approaches to DM coupling in 3$d$ oxides concern a problem of allocation of the Dzyaloshinskii vector and respective "weak" (anti)ferromagnetic moments, and  full neglect of spin-orbital effects for "nonmagnetic" oxygen O$^{2-}$ ions, which are usually believed to play only  indirect intervening role. From the other hand, the oxygen $^{17}$O NMR-NQR studies of weak ferromagnet La$_2$CuO$_4$\cite{Walstedt,PRB-2007}  seem  to evidence unconventional local oxygen "weak-ferromagnetic" polarization whose origin cannot be explained in frames of current models.

 All the systems  described above were somehow or other connected with insulating weak ferromagnets where DM coupling manifests itself in the canting of a basic antiferromagnetic structure. However, DM coupling can induce  helimagnetic distortion of the ferromagnetic order as in caesium cupric chloride, CsCuCl$_3$ to be a unique screw antiferroelectric crystal (see, e.g., Ref.\,\cite{CM-2019} and references therein).
In fact, it is known for a long time that the DM coupling can produce long-period magnetic spiral structures in ferromagnetic and antiferromagnetic crystals lacking inversion symmetry. This effect was suggested for metallic MnSi and other crystals with B20 structure, and it has been carefully proved that the sign of the DM coupling, hence the sign of the spin helix, is determined by the crystal handedness.

Phenomenologically antisymmetric DM coupling in a continual approximation  gives rise to so-called Lifshitz invariants, energy contributions linear in first spatial derivatives of the magnetization ${\bf m}({\bf r})$\,\cite{Lifshits}
\begin{equation}
	m_i\frac{\partial m_j}{\partial x_l}-m_j\frac{\partial m_i}{\partial x_l}
\end{equation}
($x_l$ is a spatial coordinate). These chiral interactions derived from the DM coupling stabilize localized (vortices) and spatially modulated structures with a fixed rotation sense of the magnetization\,\cite{Bogdanov}. In fact, these are the only mechanism to induce nanosize skyrmion structures in condensed matter.
Despite a clear weakness of the typical DM coupling as compared with typical isotropic exchange interactions the DM coupling  can be a central ingredient in the stabilization of complex magnetic textures.

The DM coupling contribution to a micromagnetic free energy density $F({\bf r})$ is usually represented as follows
\begin{equation}
	F({\bf r})=\sum_i{\bf D}_i({\bf m}({\bf r}))\cdot\left[ {\bf m}({\bf r})\times \frac{\partial{\bf m}({\bf r})}{\partial r_i}\right] \, ,
	\end{equation}
where the Dzyaloshinskii vectors ${\bf D}_i({\bf m}({\bf r}))$ are considered generally to depend on magnetization direction ${\bf m}({\bf r})$\,\cite{Blugel}. Within {\it ab initio} density functional theory (DFT) methods, DM coupling is often computed by adding spin-orbital interaction perturbatively to spirals with finite wavevectors ${\bf q}$ and extracting $D_j$ from the $q$-linear term in the dispersion $E({\bf q})$.
 However, nearly exclusively, theoretical studies in this context were in the past bound to pure spin models without itinerancy, leaving the impact of charge fluctuations aside.

 In recent years interest has shifted towards other manifestation of the DM coupling, such as the magnetoelectric effect\,\cite{KNB,Dagotto,SLD}, where reliable theoretical predictions have been lacking.
All this stimulated the critical revisit of many old  approaches to the spin-orbital effects in 3$d$ oxides, starting from the choice of proper perturbation scheme and effective spin Hamiltonian model, implied usually  only  indirect intervening role played by "nonmagnetic"\, oxygen O$^{2-}$ ions.

In this paper we present an  overview of the microscopic theory of the DM coupling and
 other related exchange-relativistic effects focusing on orthoferrites RFeO$_3$.
The rest part of the paper is organized as follows. In Sec.\,2 we shortly address main results of the microscopic theory of the isotropic superexchange interactions for so-called $S$-type ions focusing on the angular dependence of the exchange integrals. Sec.\,3 is devoted to microscopic theory of the DM coupling. Starting with Moriya's theory we arrive at a more comprehensive derivation of the Dzyaloshinskii vector for the $S$-type 3$d$-ions, its value, orientation, and sense (sign)  under different types of the (super)exchange interaction and crystal field. Here we consider the
DM coupling with participation of rare-earth ions.
Theoretical predictions of this section are compared in Sec.\,4 with experimental data for the overt and hidden canting, as well as magnetic anisotropy in orthoferrites. In Sec.\,5 we address unconventional properties of weak ferrimagnetism as a novel type of magnetic ordering in systems with competing signs of the Dzyaloshinskii vector, in particular, features of the 4$f$\,-\,3$d$ interaction in weak ferrimagnets RFe$_{1-x}$Cr$_x$O$_3$, and unconventional spin-reorientation transitions in weak ferrimagnets.
In Sec.\,6 we discuss several experimental tools to examine the sign of the Dzyaloshinskii vector, including $\mu$SR of positive muons and ligand NMR in weak ferromagnets.
Last part of the article is devoted to related exchange-relativistic effects, in particular, to
exchange-relativistic two-ion  anisotropy (Sec.\,7), antisymmetric supertransferred hyperfine interaction as electron-nuclear counterpart of DM coupling (Sec.\,8), antisymmetric exchange-relativistic spin-other-orbit coupling determining unconventional magnetooptics of weak ferromagnets (Sec.\,9), and antisymmetric exchange-relativistic spin-dependent electric polarization (Sec.\,10). A brief conclusion is made in Sec.\,11.

\section{Microscopic theory of the isotropic superexchange coupling}

DM coupling is derived from the off-diagonal (super)exchange coupling and does usually accompany a conventional (diagonal) Heisenberg type isotropic (super)exchange coupling:
\begin{equation}
	{\hat V}_{ex}=J_{12}({\bf S}_1\cdot{\bf S}_2)	\, .
\end{equation}
 The modern microscopic theory of the (super)exchange coupling had been elaborated by many physicists starting with well-known papers by P. Anderson\,\cite{PWA}, especially  intensively in 1960-70th (see review articles\,\cite{superexchange}). Numerous papers devoted to the problem pointed to existence  of many hardly estimated exchange mechanisms, seemingly comparable in value, in particular, for superexchange via intermediate ligand ion to be the most interesting for strongly correlated systems such as 3$d$ oxides. Unfortunately, up to now we have no reliable estimations of the exchange parameters, though from the other hand we have no reliable experimental information about their magnitudes. To that end, many efforts were focused on the fundamental points such as many-electron theory and orbital dependence\,\cite{1968,Levy,1971,Veltrusky}, crystal-field effects\,\cite{Sidorov}, off-diagonal exchange\,\cite{1970}, exchange in excited states\,\cite{Luk}, angular dependence of the superexchange coupling\,\cite{1971}. The irreducible tensor operators (the Racah algebra) were shown to be very instructive tool both for description and  analysis of the exchange coupling in the 3$d$- and 4$f$-systems\,\cite{1968,Levy,1971,Veltrusky,Sidorov}.

First poor man's microscopic derivation for the  dependence of the superexchange integral on the bonding angle (see Fig.\,\ref{fig1}) was performed by the author in 1970\,\cite{1971}  under simplified assumptions. For the $S$-ions with configuration 3$d^5$ (Fe$^{3+}$, Mn$^{2+}$) the angular dependence of the superexchange integral is
\begin{equation}
	J_{12}(\theta )=a+b\cdot cos\theta_{12} + c\cdot cos^2\theta_{12} \, ,
	\label{angle}
	\end{equation}
where  parameters $a$, $b$, $c$ depend on the cation-ligand separation. Parameters $a$ and $c$
are related with the contribution of the intermediate ligand 2$p$ shell,   while the $\propto\cos\theta_{12}$ term is related with the low-energy ligand inter-configurational  $2p\rightarrow 3s$ excitations.

Later on the derivation had been generalized for the 3$d$ ions in a strong cubic crystal field\,\cite{thesis}.
Orbitally isotropic contribution to the exchange integral for pair of 3$d$-ions with configurations $t_{2g}^{n_1}e_g^{n_2}$ can be written as follows
\begin{equation}
J=\sum_{\gamma_i,\gamma_j}J(\gamma_i\gamma_j)\,(g_{\gamma_i}-1)\, (g_{\gamma_j}-1)	\, ,
\end{equation}
where $g_{\gamma_i}, g_{\gamma_j}$ are effective "$g$-factors" of the $\gamma_i, \gamma_j$ subshells of ion 1 and 2, respectively:
\begin{equation}
	g_{\gamma_i} =1+\frac{S(S+1)+S_i(S_i+1)-S_j(S_j+1)}{2S(S+1)}\,  .
\end{equation}
 Kinetic exchange contribution to partial exchange parameters $I(\gamma_i\gamma_j)$ related with the electron transfer to partially filled shells can be written as follows\,\cite{Sidorov,thesis}
$$
J(e_ge_g)=\frac{(t_{ss}+t_{\sigma\sigma}cos\theta)^2}{2U};\,
J(e_gt_{2g})=\frac{t_{\sigma\pi}^2}{3U}sin^2\theta;
$$
\begin{equation}
J(t_{2g}t_{2g})=\frac{2t_{\pi\pi}^2}{9U}(2-sin^2\theta )\, ,
\label{kinetic}	
\end{equation}
where $t_{\sigma\sigma}>t_{\pi\sigma}>t_{\pi\pi}>t_{ss}$ are positive definite $d$--$d$ transfer integrals, $U$ is a mean $d$--$d$ transfer energy (correlation energy).
All the partial exchange integrals appear to be positive or "antiferromagnetic", irrespective of the bonding angle value,  though the combined effect of the $ss$ and $\sigma\sigma$ bonds $\propto cos\theta$ in $J(e_ge_g)$ yields a ferromagnetic contribution given bonding angles $\pi /2<\theta <\pi$. It should be noted that the "large" ferromagnetic potential contribution\,\cite{Freeman} has a similar angular dependence\,\cite{Luk}.
\begin{figure}[t]
\centering
\includegraphics[width=7.5cm,angle=0]{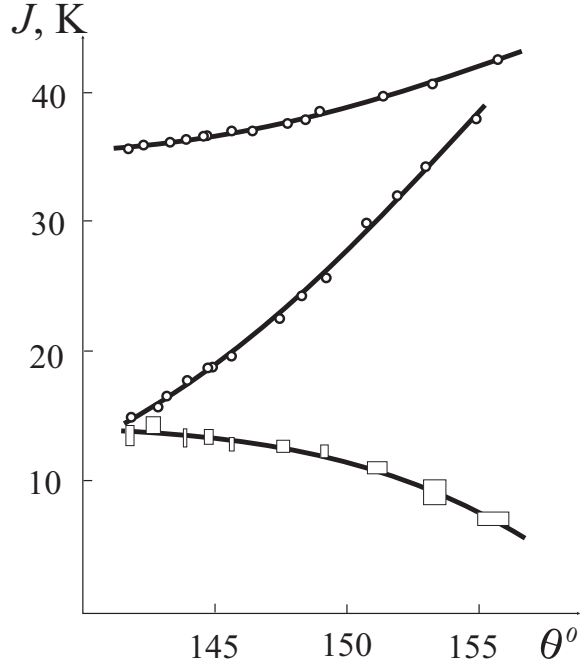}
\caption{Dependence of the Fe$^{3+}$\,-\,Fe$^{3+}$, Cr$^{3+}$\,-\,Cr$^{3+}$, Fe$^{3+}$\,-\,Cr$^{3+}$ exchange integrals on the superexchange bond angle in orthoferrites-orthocromites\,\cite{Ovanesyan}.}
\label{fig2}
\end{figure}

Some predictions regarding the relative magnitude of the $J(\gamma_i\gamma_j)$ exchange  parameters can be made using the relation among different $d$--$d$ transfer integrals as follows
\begin{equation}
	t_{\sigma\sigma}\,:\,t_{\pi\sigma}\,:\,t_{\pi\pi}\,:\,t_{ss}\approx \lambda_{\sigma}^2\,:\,\lambda_{\pi}\lambda_{\sigma}\,:\,\lambda_{\pi}^2\,:\,\lambda_{s}^2 \, ,
	\label{t-lambda}
\end{equation}
where $\lambda_{\sigma}, \lambda_{\pi}, \lambda_{s}$ are covalency parameters.
The simplified kinetic exchange contribution (\ref{kinetic}) related with the electron transfer to partially filled shells does not account for intra-center correlations which are of a particular importance for the  contribution related with the electron transfer to empty shells. For instance, appropriate contributions related with the transfer to empty $e_g$ subshell for the Cr$^{3+}$\,-\,Cr$^{3+}$ and Fe$^{3+}$\,-\,Cr$^{3+}$ exchange integrals are
$$
\Delta J_{CrCr}=-\frac{\Delta E(35)}{6U}\frac{t_{\sigma\pi}^2}{U}sin^2\theta \, ;
$$
\begin{equation}
	\Delta J_{FeCr}=-\frac{\Delta E(35)}{10U}\left[\frac{(t_{ss}+t_{\sigma\sigma}cos\theta)^2}{U}+\frac{t_{\sigma\pi}^2}{U}sin^2\theta\right] \, ,
\end{equation}
where $\Delta E(35)$ is the energy separation between $^3E_g$ and $^5E_g$ terms for $t_{2g}^3e_g$ configuration (Cr$^{2+}$ ion). Obviously these contributions have a ferromagnetic sign. Furthermore, the exchange integral $J(CrCr)$ can change sign  at $\theta$\,=\,$\theta_{cr}$:
\begin{equation}
	sin^2\theta_{cr}=\frac{1}{\left(\frac{1}{2}+\frac{3}{8}\frac{\Delta E(35)}{U}\frac{t^2_{\sigma\pi}}{t^2_{\pi\pi}}\right)} \, .
\end{equation}
Microscopically derived angular dependence of the superexchange integrals does nicely describe the experimental data for exchange integrals $J(FeFe)$, $J(CrCr)$, and $J(FeCr)$ in orthoferrites, orthochromites, and orthoferrites-orthochromites\,\cite{Ovanesyan} (see Fig.\,\ref{fig2}). The fitting allows us to predict the sign change for  $J(CrCr)$ and $J(FeCr)$ at $\theta_{12}$\,$\approx$\,133$^{\circ}$ and 170$^{\circ}$, respectively. In other words, the Cr$^{3+}$\,-\,O$^{2-}$\,-\,Cr$^{3+}$ (Fe$^{3+}$\,-\,O$^{2-}$\,-\,Cr$^{3+}$) superexchange coupling becomes ferromagnetic at $\theta_{12}\leq 133^{\circ}$ ($\theta_{12}\geq 170^{\circ}$).
However, it should be noted that too narrow (141$^{\circ}$--\,156$^{\circ}$) range of the superexchange bonding angles we used for the fitting with  assumption of the same Fe(Cr)-O bond separations and mean superexchange bonding angles for  all the systems gives rise to a sizeable parameter's uncertainty, in particular, for $J(FeFe)$ and $J(FeCr)$. In addition it is necessary to note a large uncertainty regarding what is here called the "experimental"\, value of the exchange integral. The fact is that the "experimental"\, exchange integrals we have just used above are calculated using simple MFA (mean-field approximation) relation
\begin{equation}
	T_N=\frac{zS(S+1)}{3k_B}J \, ,
\end{equation}
however, this relation yields the exchange integrals ($J$\,=\,36.8\,K in YFeO$_3$) that can be one and a half or even twice less than the values obtained by other methods\,\cite{thesis,LuCrO3,YFeO3-2014}. Most recent precise experimental data yield for YFeO$_3$ $J_c$\,=\,58.2\,K, $J_{ab}$\,=\,53.6\,K\,\cite{Park} or $J_c$\,=\,$J_{ab}$\,=\,51.5\,K\,\cite{Amelin}.

Above we addressed only typically antiferromagnetic kinetic (super)exchange contribution as a result of the second order perturbation theory. However, actually this contribution does compete with typically ferromagnetic potential (super)exchange contribution, or Heisenberg exchange, which is a result of the first order perturbation theory. The most important contribution to the potential superexchange can be related with the intra-atomic ferromagnetic Hund exchange interaction of unpaired electrons on orthogonal ligand orbitals hybridized with the $d$-orbitals of the two nearest magnetic cations.

Strong dependence of the $d$--$d$ superexchange integrals  on the cation-ligand-cation separation is usually described by the Bloch's rule\,\cite{Bloch}:
\begin{equation}
	\frac{\partial\ln J}{\partial\ln R}=\frac{\partial J}{\partial R}/\frac{J}{R} \approx -\,10 \, .
\end{equation}

\section{Microscopic theory of the DM coupling}

\subsection{Moriya's theory}

First microscopic theory of weak ferromagnetism, or theory of anisotropic superexchange interaction was provided by Moriya\,\cite{Moriya} who extended the Anderson theory of superexchange to include spin-orbital coupling $V_{so}=\sum_i \xi({\bf l}_i\cdot{\bf s}_i)$.
Moriya started with the one-electron Hamiltonian for $d$-electrons as follows
$$
	{\hat H}=\sum_{fm\sigma}\epsilon_m{\hat d}^{\dagger}_{fm\sigma}{\hat d}_{fm\sigma}+\sum_{m\not= m^{\prime}, \sigma}t_{fmf^{\prime}m^{\prime}}{\hat d}^{\dagger}_{fm\sigma}{\hat d}_{f^{\prime}m^{\prime}\sigma}+
$$
\begin{equation}
\sum_{fm\not= f^{\prime}m^{\prime}, \sigma\sigma^{\prime}}{\hat d}^{\dagger}_{fm\sigma}({\bf C}_{fmf^{\prime}m^{\prime}}\cdot{\pmb \sigma}){\hat d}_{f^{\prime}m^{\prime}\sigma^{\prime}} \, ,
\end{equation}
where
\begin{equation}
{\bf C}_{fmf^{\prime}m^{\prime}}=-\frac{\xi}{2}\sum_{m^{\prime\prime}}\left(\frac{{\bf l}_{fmfm^{\prime\prime}}t_{fm^{\prime\prime}f^{\prime}m^{\prime}}}{\epsilon_{m^{\prime\prime}}-\epsilon_m}+\frac{t_{fmf^{\prime}m^{\prime\prime}}{\bf l}_{f^{\prime}m^{\prime\prime}f^{\prime}m^{\prime}}}{\epsilon_{m^{\prime\prime}}-\epsilon_{m^{\prime}}}\right)	
\label{C}
\end{equation}
is a spin-orbital correction to transfer integral. Then Moriya did calculate the generalized Anderson kinetic exchange that contains both conventional isotropic exchange and anisotropic symmetric and antisymmetric terms, that is quasidipole anisotropy and DM coupling, respectively.
The expression for the Dzyaloshinskii vector
\begin{equation}
	{\bf d}_{ff^{\prime}}=\frac{4i}{U}\sum_{m\not= m^{\prime}}\left[t_{fmf^{\prime}m^{\prime}}{\bf C}_{f^{\prime}m^{\prime}fm}-{\bf C}_{fmf^{\prime}m^{\prime}}t_{f^{\prime}m^{\prime}fm}\right] \, .
\end{equation}
has been obtained by Moriya assuming orbitally nondegenerate ground states $m$ and $m^{\prime}$ on sites $f$ and $f^{\prime}$, respectively.
 It is worth noting that the spin-operator form of the DM coupling  follows from the relation:
\begin{equation}
	{\bf S}_1({\bf S}_1\cdot{\bf S}_2)+ ({\bf S}_1\cdot{\bf S}_2) {\bf S}_2=-i[{\bf S}_1\times{\bf S}_2]\, .
\end{equation}
Moriya found the symmetry constraints on the orientation of the Dzyaloshinskii vector $\mathbf{d}_{ij}$.
Let two ions 1 and 2 are located at the points A and B, respectively, with C point bisecting the AB line:

1. When C is a center of inversion: D=0.

2. When a mirror plane $\perp$AB passes through C,
     ${\bf D} \parallel$ mirror plane or  ${\bf D} \perp$AB.

3. When there is a mirror plane including A and B,
     ${\bf D} \perp$ mirror plane.

4. When a twofold rotation axis $\perp$AB passes through C,
    ${\bf D} \perp$ twofold axis.

5. When there is an n-fold axis (n$
\geq$2) along AB,  ${\bf D} \parallel$AB.


Despite its seeming simplicity the operator form of the DM coupling (\ref{DM})  raises some questions and doubts. First, at variance with the scalar product  $\left({\bf S}_1\cdot{\bf S}_2\right)$  the vector product of the  spin operators $\left[{\bf S}_1\times{\bf S}_2\right]$ changes the spin multiplicity, that is the net spin ${\bf S}_{12}={\bf S}_1+{\bf S}_2$, that underscores the need for quantum description. Spin nondiagonality of the DM coupling implies very unusual features of the ${\bf D}$-vector somewhat resembling vector orbital operator whose transformational properties cannot be isolated from the lattice\,\cite{Dmitrienko-2010}. It seems  the ${\bf D}$-vector  does not transform as a vector at all.

Another issue that causes some concern is the structure and location of the ${\bf D}$ vector and corresponding spin cantings. Obviously, the ${\bf D}_{12}$ vector should be related in one or another way to spin-orbital contributions localized on sites 1 and 2, respectively. These components  may differ in their magnitude and direction, while the operator form (\ref{DM}) implies some  averaging both for ${\bf D}_{12}$ vector and spin canting between the two sites.

Moriya did not take into account the effects of the crystal field symmetry and strength and did not specify the character of the (super)exchange coupling, that, as we'll see below, can crucially affect the direction and value of the Dzyaloshinskii vector up to its vanishing. Furthermore, he made use of a very simplified form (\ref{C}) of the spin-orbital perturbation correction to  the transfer integral (see Exp.\,(2.5) in Ref.\,\cite{Moriya}). The fact is that the structure of the charge transfer matrix elements implies the  involvement of several different on-site configurations ($t_{kn}\propto \langle N_1-1N_2+1|{\hat H}|N_1N_2\rangle$). Hence, the  perturbation correction has to be more complicated than (\ref{C}), at least, it should involve the spin-orbital matrix elements (and excitation energies!) for one- and two-particle configurations. As a result, it does invalidate the author's conclusion about the equivalence of the two perturbation schemes, based on the $V_{SO}$ corrections to the transfer integral and to the exchange coupling, respectively.
Another limitation of the Moriya's theory is related to a full neglect of the ligand spin-orbital contribution to DM coupling.
Despite these shortcomings the Moriya's estimation for the ratio between the  magnitudes of the Dzyaloshinskii vector $d=|{\bf d}|$ and isotropic exchange $J$: $d/J\approx \Delta g/g$, where $g$ is the gyromagnetic ratio, $\Delta g$ is its deviation from the free-electron value, respectively, in some cases may be helpful, however, only for a very  rough estimation.

\subsection{Microscopic theory of the DM coupling: superexchange interaction of the $S$-ions}

The first microscopic theory of the DM coupling for the superexchange bond M$_1$--O--M$_2$ proposed by the author\,\cite{1971} (see also Refs.\,\cite{thesis,1977,CM}) assumed the interaction of "free" ions with
the ground $^6S$ state of the 3$d^5$ configuration  (Mn$^{2+}$, Fe$^{3+}$), interacting through an intermediate anion of the O$^{2-}$ type.
Final expression for the Dzyaloshinskii vector was obtained as follows
\begin{equation}
	\mathbf{d}_{12} = d_{12}(\theta_{12}) [\mathbf{r}_1 \times \mathbf{r}_2] \, ,
	\label{d}
\end{equation}
with
\begin{equation}
	d_{12}(\theta)=d_1(R_{10},R_{20})+d_2(R_{10},R_{20})cos\theta_{12} \, ,
	\label{dcos}
\end{equation}
where the first and the second terms are determined by the superexchange mechanisms related with the ligand inter-configurational  2$p\rightarrow 3s$ excitations and intra-configurational 2$p$--2$p$ effects, respectively. It should be noted that given $\theta$\,=\,$\theta_{cr}$, where
\begin{equation}
	cos\,\theta_{cr}=-d_1/d_2
\end{equation}
the   Dzyaloshinskii vector changes its sign.

However, later it was shown\,\cite{thesis,1977,CM} that the correct theory of the Dzyaloshinskii interaction even for $S$-type ions should take into account the crystal field effects.

As the most illustrative example we consider a pair of 3$d^5$ ions such as Fe$^{3+}$, or Mn$^{2+}$ with the ground state $^6S$ in an $intermediate$ octahedral crystal field which does split the $^{2S+1}L$ terms into crystal  $^{2S+1}L\Gamma$ terms and mix the crystal terms with the same octahedral symmetry, that is with the same $\Gamma$'s\,\cite{STK}. Spin-orbital coupling does mix the $^6S$ ground state with the $^4PT_{1g}$ term,
however the $^4PT_{1g}$ term has been mixed with other $^4T_{1g}$ terms,  $^4FT_{1g}$ and  $^4GT_{1g}$. Namely the latter effect appears to be  a decisive factor for appearance of the DM coupling. The $|4(L)T_{1g}\rangle $ wave functions can be easily calculated by a standard technique\,\cite{STK} as follows\,\cite{thesis}:
$$
|4(P)T_{1g}\rangle = 0.679 |4PT_{1g}\rangle -0.604 |4FT_{1g}\rangle  +0.418 |4GT_{1g}\rangle \, ;
$$
$$
|4(F)T_{1g}\rangle = 0.387 |4PT_{1g}\rangle +0.777 |4FT_{1g}\rangle  +0.495 |4GT_{1g}\rangle \, ;
$$
\begin{equation}
|4(G)T_{1g}\rangle = -0.604 |4PT_{1g}\rangle -0.169 |4FT_{1g}\rangle  +0.737 |4GT_{1g}\rangle \, ,
\label{4T1} 	
\end{equation}
given the crystal field and intra-atomic correlation parameters\,\cite{STK} typical for orthoferrites\,\cite{Kahn}: 10Dq\,=\,12200\,$cm^{-1}$; B\,=\,700\,$cm^{-1}$; C\,=\,2600\,$cm^{-1}$. We see that due to the crystal-field mixing effect all the three crystal terms $^4PT_{1g}$, $^4FT_{1g}$, and $^4CT_{1g}$ will contribute to the DM coupling. Furthermore, the overall nonzero contribution will be determined by the ${}^4P-{}^4G$ mixing\,\cite{CM}.
\begin{table*}
\begin{center}
\caption{Wave functions and energies for the $^4$T$_{1g}$ terms for Fe$^{3+}$ ion in orthoferrites}
\begin{tabular}[t]{|c|l|}
\hline
 Wave function & Energy, cm$^{-1}$ \\
\hline
$|{}^4T_{1g}(41)\rangle = 0.988 |t_{2g}^{4}e_g^{1}{}^4T_{1g}\rangle -0.123 |t_{2g}^{3}e_g^{2}{}^4T_{1g}\rangle  +0.088 |t_{2g}^{2}e_g^{3}{}^4T_{1g}\rangle$ & $E(41)= 0.96\cdot 10^4$ \\
\hline
$|{}^4T_{1g}(32)\rangle = 0.058 |t_{2g}^{4}e_g^{1}{}^4T_{1g}\rangle +0.844 |t_{2g}^{3}e_g^{2}{}^4T_{1g}\rangle  -0.534 |t_{2g}^{2}e_g^{3}{}^4T_{1g}\rangle$  & $E(32)= 2.96\cdot 10^4$ \\
\hline
$|{}^4T_{1g}(23)\rangle = -0.140 |t_{2g}^{4}e_g^{1}{}^4T_{1g}\rangle -0.522 |t_{2g}^{3}e_g^{2}{}^4T_{1g}\rangle  +0.841 |t_{2g}^{2}e_g^{3}{}^4T_{1g}\rangle$ & $E(23)= 3.69\cdot 10^4$ \\
\hline
\end{tabular}
\end{center}
\label{table4T}
\end{table*}
However, it is more appropriate to consider the interaction of 3$d$ ions in a $strong$ crystal field scheme.
Hereafter we address the DM coupling for the  $S$-type magnetic 3$d$ ions with orbitally nondegenerate high-spin ground state in a strong cubic crystal field, that is for the 3$d$ ions with half-filled shells $t_{2g}^3$, $t_{2g}^3e_g^2$, $t_{2g}^6e_g^2$ and ground states $^4A_{2g}$,  $^6A_{1g}$, $^3A_{2g}$, respectively\,\cite{thesis,1977,CM}.
In particular, for the ${}^{4}T_{1g}$ terms of the 3$d^5$ ion in the strong cubic crystal field approximation instead of expressions (\ref{4T1}) we arrive at the wave functions of the $t_{2g}^{n_1}e_g^{n_2}$ configurations as shown in Table\,2.


Making use of expressions for spin-orbital coupling $V_{SO}$ and main kinetic contribution to the  superexchange parameters, that define DM coupling after routine algebra we have found that the DM coupling can be written in a standard form (\ref{d}), where  $d_{12}$ can   be written as follows\,\cite{1977,thesis,CM}
\begin{equation}
	d_{12}=X_1Y_2+X_2Y_1 \, ,
	\label{XY}
\end{equation}
where the $X$ and $Y$ factors do reflect the exchange-relativistic structure of the second-order perturbation theory and details of the electron configuration for $S$-type ion. The exchange factors $X$ are
\begin{equation}
	X_i=\frac{(g^{(i)}_{e_g}-1)}{2U}t_{\pi\sigma}(t_{ss}+t_{\sigma\sigma}cos\theta )-\frac{(g^{(i)}_{t_{2g}}-1)}{3U}t_{\pi\pi}t_{\sigma\pi}cos\theta  \, ,
	\label{X}
\end{equation}
 where $g^{(i)}_{e_g}$, $g^{(i)}_{t_{2g}}$ are effective $g$-factors for $e_g$, ${t_{2g}}$ subshells, respectively, $t_{\sigma\sigma}>t_{\pi\sigma}>t_{\pi\pi}>t_{ss}$ are positive definite $d$--$d$ transfer integrals, $U$ is a $d$--$d$ transfer energy (correlation energy). The general form of the dimensionless factors $Y$ determined by spin-orbital constants  and excitation energies is more complicated (see, e.g., Ref.\,\cite{CM}).
The both factors $X$ and $Y$ are presented in Table\,\ref{tableXY} for $S$-type 3$d$-ions, where  $\xi_{3d}$ is the spin-orbital parameter, $\Delta E_{^{2S+1}\Gamma}$ is the energy of the excited $^{2S+1}\Gamma$ crystal term interacting with the ground state due to $V_{SO}$.

The signs for $X$ and $Y$ factors in Table\,\ref{tableXY} are predicted for rather large superexchange bonding angles $|cos\theta_{12}|>t_{ss}/t_{\sigma\sigma}$ which are typical for many 3$d$ compounds such as oxides and a relation $\Delta E_{^4T_{1g}}(41)<\Delta E_{^4T_{1g}}(32)$ which is typical for high-spin 3$d^5$ configurations. Excited configuration $t_{2g}^2e_g^3$ does not contribute to the DM coupling for 3$d^5$ ions.

It is worth noting that earlier we have detected and corrected a casual and unintentional
error in sign of the $X_i$ parameters having made both in our earlier papers\,\cite{1977,thesis} and  recent paper Ref.\,\cite{2016}.
Hereafter we present correct signs for $X_i$ in (\ref{X}) and Table\,\ref{tableXY}\,\cite{CM}.

Rather simple expressions for the factors $X_i$ and $Y_i$ do not take into account the mixing/interaction effects for the ${}^{2S+1}\Gamma$  terms with the same symmetry and the contribution of empty subshells to the exchange coupling (see Ref.\,\cite{thesis}). Nevertheless, the data in Table\,\ref{tableXY} allow us to evaluate both the  numerical value and sign of the $d_{12}$ parameters.

It should be noted that for critical angle $\theta_{cr}$, when the Dzyaloshinskii vector changes its sign we have
$	cos\,\theta_{cr}=-d_1/d_2=\frac{\lambda_s^2}{\lambda_{\sigma}^2}$ for $d^8$\,-\,$d^8$ pairs and $	 cos\,\theta_{cr}=-d_1/d_2=\frac{\lambda_s^2}{\lambda_{\sigma}^2-\lambda_{\pi}^2}$ for $d^5$\,-\,$d^5$ pairs. Making use of different experimental data for covalency parameters (see, e.g., Ref.\,\cite{Tofield}) we arrive at $d_1/d_2\sim \frac{1}{5}-\frac{1}{3}$ and $\theta_{cr}\approx 100^{\circ}-110^{\circ}$ for Fe$^{3+}$\,-\,Fe$^{3+}$ pairs in oxides.

 Relation among different $X$'s given the superexchange geometry and covalency parameters typical for orthoferrites and orthochromites \,\cite{thesis} is
\begin{equation}
	|X_{d^8}|\geq |X_{d^3}|\geq |X_{d^5}| \, ,
\end{equation}
however, it should be underlined its sensitivity both to superexchange geometry and covalency parameters. Simple comparison of the exchange parameters $X$(see (\ref{X}) and Table \ref{tableXY}) with exchange parameters $I(\gamma_i\gamma_j)$ (\ref{kinetic}) evidences their close magnitudes. Furthermore, the relation (\ref{t-lambda}) allows us to maintain more definite correspondence.

Given typical values of the cubic crystal field parameter $10Dq\approx$\,1.5\,eV we arrive at a relation among different $Y$'s\,\cite{thesis}
\begin{equation}
	|Y_{d^8}|\geq |Y_{d^5}|\geq |Y_{d^3}|
\end{equation}
with $Y_{d^8}\approx 7.0\cdot 10^{-2}$, $Y_{d^5}\approx -2.5\cdot 10^{-2}$, $Y_{d^3}\approx 1.5\cdot 10^{-2}$.
\begin{center}
\begin{table*}
\caption{Expressions for the $X$ and $Y$ parameters that define the magnitude and the sign of the Dzyaloshinskii vector in pairs of
the $S$-type 3$d$-ions with local octahedral symmetry. Signs for $X_i$ correspond to the bonding angle $\theta >\theta_{cr}$.}
\begin{tabular}{|c|c|c|c|c|c|}
\hline
   \begin{tabular}{c}
Ground state \\
configuration \\
\end{tabular}        & $X$ & Sign $X$ &  $Y$ & Sign $Y$ & \begin{tabular}{c}
Excited state \\
configuration \\
\end{tabular}  \\ \hline
  \begin{tabular}{c}
3$d^3$($t_{2g}^3$):${}^4A_{2g}$ \\
V$^{2+}$, Cr$^{3+}$, Mn$^{4+}$ \\
\end{tabular}  & $-\frac{1}{3U}t_{\pi\pi}t_{\sigma\pi}cos\theta$ & + &$\frac{2\xi_{3d}}{3\sqrt{3}}(\frac{1}{\Delta E_{^4T_{2g}}}+\frac{2}{\Delta E_{^2T_{2g}}})$ & + & $t_{2g}^2e_g^1$ \\ \hline
 \begin{tabular}{c}
3$d^5$($t_{2g}^3e_g^2$):${}^6A_{1g}$ \\
Mn$^{2+}$, Fe$^{3+}$ \\
\end{tabular}  & \begin{tabular}{c}$-\frac{1}{5U}(t_{\pi\pi}t_{\sigma\pi}cos\theta$ - \\
$t_{\pi\sigma}\left(t_{ss}+t_{\sigma\sigma}cos\theta )\right)$\\
\end{tabular} & -- & --$\frac{6\xi_{3d}}{5\sqrt{3}}(\frac{1}{\Delta E_{^4T_{1g}}(41)}-\frac{1}{\Delta E_{^4T_{1g}}(23)})$ & -- & $t_{2g}^4e_g^1$, $t_{2g}^2e_g^3$ \\ \hline	
\begin{tabular}{c}
3$d^8$($t_{2g}^6e_g^2$):${}^3A_{2g}$ \\
Ni$^{2+}$, Cu$^{3+}$ \\
\end{tabular}  & $\frac{1}{2U}t_{\pi\sigma}(t_{ss}+t_{\sigma\sigma}cos\theta )$& -- & $\frac{3\xi_{3d}}{2\sqrt{3}}(\frac{1}{\Delta E_{^3T_{2g}}}+\frac{1}{\Delta E_{^1T_{2g}}})$ & + & $t_{2g}^5e_g^3$\\ \hline
	\end{tabular}
\label{tableXY}
\end{table*}
\end{center}

The greatest value of the $d_{12}$ factor  is predicted for $d^8$\,-\,$d^8$ pairs, while for $d^5$\,-\,$d^5$ pairs one expects  a much less (may be one order of magnitude) value. The $d_{12}$ factor for  $d^3$\,-\,$d^3$ pairs  is predicted to be somewhat above the value for $d^5$\,-\,$d^5$ pairs. For different pairs: $d_{12}$($d^3$\,-\,$d^5$)$\approx$ -$d_{12}$($d^3$\,-\,$d^3$);\,$d_{12}$($d^8$\,-\,$d^5$)$\approx$ $d_{12}$($d^5$\,-\,$d^5$); $d_{12}$($d^3$\,-\,$d^8$)$\geq$ $d_{12}$($d^3$\,-\,$d^3$). Puzzlingly, that despite strong isotropic exchange coupling for $d^5$\,-\,$d^5$ and $d^5$\,-\,$d^8$ pairs, the DM coupling for these pairs is expected to be  the least one among the $S$-type pairs. For $d^5$\,-\,$d^5$ pairs, in particular, Fe$^{3+}$\,-\,Fe$^{3+}$ we have two compensation effects.
First, the $\sigma$-bonding contribution to the $X$ parameter is partially compensated by the $\pi$-bonding contribution, second, the contribution of the ${}^4T_{1g}$ term of the $t_{2g}^4e_g^1$ configuration is partially compensated by the contribution of the ${}^4T_{1g}$ term of the $t_{2g}^2e_g^3$ configuration.

Theoretical predictions of the corrected sign of the Dzyaloshinskii vector in pairs of the $S$-type 3$d$-ions with local octahedral symmetry (the sign rules) are presented in Table\,\ref{tablesign}. The signs for $d^3$\,-\,$d^3$, $d^5$\,-\,$d^5$, and $d^3$\,-\,$d^8$ pairs turn out to be the same but opposite  to signs for $d^3$\,-\,$d^5$ and $d^8$\,-\,$d^8$ pairs.
In a similar way to how different signs of the conventional exchange integral determine different (ferro-antiferro) magnetic orders the different signs of the Dzyaloshinskii vectors  create a possibility of nonuniform  (ferro-antiferro) ordering of local weak (anti)ferromagnetic moments, or local overt/hidden cantings. Novel magnetic phenomenon and novel class of magnetic materials, which are systems such as solid solutions YFe$_{1-x}$Cr$_x$O$_3$ with competing signs of the Dzyaloshinskii vectors will be addressed below (Sec.\,5) in more detail.

\begin{table}
\begin{center}
\caption{Theoretical predictions of the sign of the Dzyaloshinskii vector in pairs of the $S$-type 3$d$-ions
 with local octahedral symmetry and the bonding angle $\theta >\theta_{cr}$.}
\begin{tabular}{|c|c|c|c|}
\hline
   3$d^n$      & 3$d^3$($t_{2g}^3$) &  3$d^5$($t_{2g}^3e_g^2$)  & 3$d^8$($t_{2g}^6e_g^2$)  \\ \hline
  3$d^3$($t_{2g}^3$)  & + &--& + \\ \hline
 3$d^5$($t_{2g}^3e_g^2$) & -- &+ & + \\ \hline	
3$d^8$($t_{2g}^6e_g^2$) & +& +& -- \\ \hline
	\end{tabular}
\label{tablesign}
\end{center}
\end{table}

\subsection{DM coupling in trigonal  hematite $\alpha$-Fe$_2$O$_3$ and borate FeBO$_3$}
Making use of our theory based on the bare ideal octahedral symmetry of $S$-type ions to the classical weak ferromagnet $\alpha$-Fe$_2$O$_3$ we arrive at a little unexpected  disappointment, as the theory does predict that the  contribution of the three equivalent Fe$^{3+}$\,-\,O$^{2-}$\,-\,Fe$^{3+}$ superexchange pathes for the two corner shared FeO$_6^{9-}$ octahedrons  to the net Dzyaloshinskii vector strictly turns into zero.  Exactly the same  result will be obtained, if we consider the direct Fe$^{3+}$\,-\,Fe$^{3+}$ exchange in the system of two ideal FeO$_6^{9-}$ octahedrons bonded through the three common oxygen ions when ${\bf R}_{12}\parallel C_3$. Obviously, it is precisely this fact that caused  a tiny spin canting in hematite being an order of magnitude smaller than, e.g., in orthoferrites RFeO$_3$ or borate FeBO$_3$. So what was the real reason of weak ferromagnetism in  $\alpha$-Fe$_2$O$_3$ as "opening a new page of weak ferromagnetism"? What is a microscopic origin of nonzero Dzyaloshinskii vector  which should be directed along the $C_3$ symmetry axis according Moriya rules?
First of all we should consider trigonal distortions for the FeO$_6^{9-}$ octahedrons which have a $T_{2}$ symmetry and give rise to a mixing of the ${}^4T_{1g}$ terms with ${}^4A_{2g}$ and ${}^4T_{2g}$ terms. The best way to solve the problem in principle is to proceed with a coordinate system where $O_z$ axis is directed along the $C_3$ symmetry axis rather than with the usually applied $O_{z}\parallel C_4$ geometry\,\cite{CM}. Symmetry analysis shows that the axial distortion along the Fe$^{3+}$\,-\,Fe$^{3+}$  bond  can induce the DM coupling with Dzyaloshinskii vector directed along the bond, however, only for locally nonequivalent Fe$^{3+}$ centers, otherwise we arrive at an exact compensation of  the contributions of the spin-orbital couplings on sites 1 and 2\,\cite{CM}.

Trigonal hematite  $\alpha$-Fe$_2$O$_3$ has the same crystal symmetry $R\overline{3}c-D_{3d}^6$ as weak ferromagnet FeBO$_3$. Furthermore, the borate  can be transformed into hematite by the Fe$^{3+}$ ion substitution for B$^{3+}$ with a displacement of both "old" and "new" iron ions along trigonal axis. As a result we arrive at  emergence of an additional strong isotropic (super)exchange coupling of three-corner-shared non-centrosymmetric FeO$_6$ octahedra with short Fe-O separations (1.942\,\AA) that determines very high N\'{e}el temperature $T_N$\,=\,948\,K in hematite as compared with $T_N$\,=\,348\,K in borate. However, the $D_{3h}$ symmetry of these exchange bonds points to a distinct compensation of the two Fe-ion's contribution to Dzyaloshinskii vector. In other words,  weak ferromagnetism in hematite $\alpha$-Fe$_2$O$_3$ is determined by the DM coupling for the  same Fe-O-Fe bonds as in borate FeBO$_3$. However, the Fe-O separations for these bonds in hematite (2.111\,\AA) are markedly longer than in borate (2.028\,\AA) that points to a significantly weaker DM coupling. Combination of the weaker DM coupling and stronger isotropic exchange in $\alpha$-Fe$_2$O$_3$ as compared with FeBO$_3$ does explain the one order of magnitude difference in canting angles.


\begin{table*}
\begin{center}
\caption{The structural factors $\left[{\bf r}_1\times{\bf r}_2\right]_{x,y,z}$ for
the superexchange coupled Fe$_1$\,-\,O\,-\,Fe$_{2,4}$ pairs in orthoferrites with numerical values for YFeO$_3$, $a, b, c$ are lattice parameters,
$l$ is a mean Fe\,-\,O separation, $x_1, y_1$, $x_2, y_2, z_2$ are oxygen parameters for the O(4c) and O(8d) positions, respectively. }
\begin{tabular}{|c|c|c|c|}
\hline
         & $\left[{\bf r}_1\times{\bf r}_2\right]_x$ &  $\left[{\bf r}_1\times{\bf r}_2\right]_y$  & $\left[{\bf r}_1\times{\bf r}_2\right]_z$  \\ \hline
  2 & $\frac{(\frac{1}{2}-y_1)bc}{2l^2}$\,=\,0.20 &-$\frac{x_1ac}{2l^2}$\,=\,-0.55 & 0 \\ \hline
  4  & $\pm\frac{z_2bc}{2l^2}$\,=\,$\pm$\,0.31 & $-\frac{z_2ac}{2l^2}$\,=\,-0.29 & $\frac{(y_2-x_2+\frac{1}{2})ab}{2l^2}$\,=\,0.41 \\ \hline
 	\end{tabular}
\label{tabler1r2}
\end{center}
\end{table*}

\subsection{DM coupling with participation of rare-earth ions}

Spin-orbital interaction for the rare-earth ions with valent $4f^n$ configuration is diagonalized within the $(LS)J$ multiplets hence the conventional DM coupling
$$
	{\hat H}_{DM}^{ff}=\sum_{m>n}({\bf d}_{mn}\cdot[{\bf S}_m\times{\bf S}_n])=
$$
 \begin{equation}
\sum_{m>n}(g_m-1)(g_n-1)({\bf d}_{mn}\cdot[{\bf J}_m\times{\bf J}_n])
\end{equation}
($g_{m,n}$ are the Lande factors) can arise for the $f$--$f$ superexchange only due to a spin-orbital contribution on intermediate ligands. Obviously, for the rare-earth-3$d$-ion (super)exchange we have an additional contribution of the 3$d$-ion spin-orbital interaction.
The rare-earth-3$d$-ion  DM coupling R$^{3+}$\,-\,O$^{2-}$\,-\,Fe$^{3+}$ (R=Nd, Gd)
\begin{equation}
	{\hat H}_{DM}^{fd}=\sum_{m>n}({\bf d}_{mn}\cdot[{\bf J}_m\times{\bf S}_n])
\end{equation}
 has been  theoretically and experimentally considered in Refs.\,\cite{Belov,GdFeCrO3}.

 Effective field at the R ion can be written as a sum of ferro- and antiferromagnetic
contributions:
\begin{equation}
	{\bf H}_{fd}=\alpha {\bf F}+{\bf \stackrel{\leftrightarrow}{\beta}}{\bf G}\, ,
\end{equation}
where $\alpha$ determines the contribution of the isotropic $f$\,-\,$d$
exchange, while tensor ${\bf \stackrel{\leftrightarrow}{\beta}}$ does the contribution of symmetric and antisymmetric anisotropic $f$\,-\,$d$ interactions. These interactions were studied in  GdFeO$_3$\,\cite{Belov}, and the authors found that
$$
	\alpha F_z= -0.19\,;
$$
\begin{equation}
\beta_{zx}G_x=(\beta_{zx}^s+\beta_{zx}^a)G_x=-0.05+0.21=0.16\,,
\label{Fe}
\end{equation}
(in Tesla). Quite unexpected, the antisymmetric antiferromagnetic contribution $\beta_{zx}^aG_x$ to effective field on Gd$^{3+}$ ion in $\Gamma_4$ magnetic phase at T\,=\,0\,K, which is determined by the $f$\,-\,$d$ DM  coupling  appears to be the leading one. Moreover, according to the data from Ref.\,\cite{GdFeCrO3}, in GdCrO$_3$
$$
	\alpha F_z= -0.13\,;
$$
\begin{equation}
\beta_{zx}G_x=(\beta_{zx}^s+\beta_{zx}^a)G_x=+0.05-0.47=-0.42\,
\label{Cr}
\end{equation}
that is the $f$\,-\,$d$ DM  coupling has value greater than in the GdFeO$_3$, however, with opposite sign, which is consistent with the different sign of the factor $Y$ for Fe$^{3+}$ and Cr$^{3+}$ (see Table\,\ref{tableXY}).

\begin{figure}[t]
\begin{center}
\includegraphics[width=8cm,angle=0]{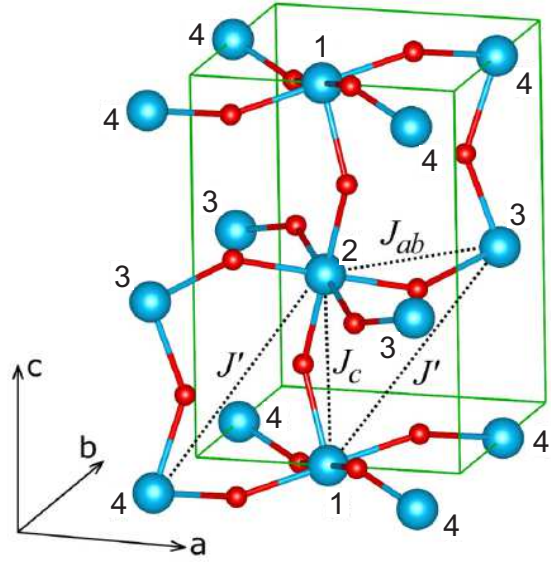}
\caption{(Color online) Structure of the Fe$^{3+}$--O$^{2-}$--Fe$^{3+}$ superexchange bonding in orthoferrites. $J_{ab}$ and $J_{c}$ are nearest-neighbor superexchange integrals, $J^{\prime}$ superexchange integral for next-nearest-neighbors. 1, 2, 3, 4, are Fe$^{3+}$ ions in four nonequivalent positions. Reproduced from Ref.\,\cite{Amelin}. }
\label{fig3}
\end{center}
\end{figure}

\section{Theoretical predictions as compared with experiment}

\subsection{Dzyaloshinskii interaction in orthoferrites}

Four Fe$^{3+}$ ions occupy positions 4b in the orthorhombic elementary cell of orthoferrites RFeO$_3$ (space group $Pbnm$):
$$
1\,(1/2, 0, 0); 2\,(1/2, 0, 1/2); 3\,(0, 1/2, 1/2); 4\,(0, 1/2, 0)\,   .
$$
Classical basis vectors of magnetic structure for 3$d$ sublattice are defined as follows:
$$
4S{\bf F}={\bf S}_1+{\bf S}_2+{\bf S}_3+{\bf S}_4\,; \\
$$
$$
4S{\bf G}={\bf S}_1-{\bf S}_2+{\bf S}_3-{\bf S}_4\,; \\
$$
$$
4S{\bf C}={\bf S}_1+{\bf S}_2-{\bf S}_3-{\bf S}_4\,; \\
$$
\begin{equation}
4S{\bf A}={\bf S}_1-{\bf S}_2-{\bf S}_3+{\bf S}_4\,.
\end{equation}
Here ${\bf G}$ describes the main antiferromagnetic component, ${\bf F}$ gives the weak ferromagnetic moment (overt canting),  the weak antiferromagnetic
components ${\bf C}$ and ${\bf A}$ describe a canting without  net magnetic moment (hidden canting). Allowed spin configurations for 3$d$-sublattice are denoted as $\Gamma_1\,(A_x, G_y, C_z)$,
$\Gamma_2\,(F_x, C_y, G_z)$, $\Gamma_4\,(G_x, A_y, F_z)$, where the components given in
parentheses are the only ones different from zero.
It is worth noting that another labeling
of the Fe$^{3+}$  positions than what was used here is found in the literature (see, for example, Refs.\,\cite{muon,Amelin}), in which case the basis vectors ${\bf G}$, ${\bf C}$, ${\bf A}$
may differ in sign.

Within simplest classical approximation the operator of symmetric and antisymmetric $d$\,-\,$d$ exchange interactions in orthoferrites
\begin{equation}
{\hat H}=\sum_{i>j}J_{ij}({\bf S}_i\cdot{\bf S}_j)+\sum_{i>j}({\bf d}_{ij}\cdot[{\bf S}_i\times{\bf S}_j])
\end{equation}
can be written in terms of basis vectors as free energy as follows (see, e.g., Ref.\,\cite{thesis} and references therein)
$$
\Phi= \frac{J_F}{2}{\bf F}^2+\frac{J_G}{2}{\bf G}^2+\frac{J_C}{2}{\bf C}^2+\frac{J_A}{2}{\bf A}^2+
$$
$$
D_x(C_yG_z-C_zG_y)+D_y(F_zG_x-F_xG_z)+D_z(A_xG_y-A_yG_x)
$$
\begin{equation}
d_x(A_yF_z-A_zF_y)+d_y(C_zA_x-C_xA_z)+d_z(C_xF_y-C_yF_x)\, ,
\end{equation}
where for the energy per ion
$$
J_F=-J_G=S^2(2J_{ab}+J_c); J_A=-J_C=S^2(2J_{ab}-J_c);
$$
$$
D_x=-S^2\sum_2d_x(12);
$$
$$
D_y=-S^2\left(\sum_4d_y(14)+\sum_2d_y(12)\right);
$$
 \begin{equation}
D_z=-S^2\sum_4d_z(14);
\end{equation}
By minimizing the free energy under condition ${\bf F}^2+{\bf G}^2+{\bf C}^2+{\bf A}^2=1$ and $F, C, A\ll G$ we find
$$
F_z=-\frac{D_y}{J_F-J_G}G_x;\,\, A_y=\frac{D_z}{J_A-J_G}G_x;
$$
$$
F_x=\frac{D_y}{J_F-J_G}G_z;\,\, C_y=-\frac{D_x}{J_C-J_G}G_z;
$$
\begin{equation}
A_x=-\frac{D_z}{J_A-J_G}G_y;\,\, C_z=\frac{D_x}{J_C-J_G}G_y;
\end{equation}

\subsection{Overt and hidden canting in orthoferrites}

Figure\,\ref{fig3} shows the intricate  structure
 of the Fe$^{3+}$--O$^{2-}$--Fe$^{3+}$ superexchange bondings in orthoferrites that points to a complicated structural dependence of the Dzyaloshinskii vectors.
\begin{figure*}[t]
\begin{center}
\includegraphics[width=12.5cm,angle=0]{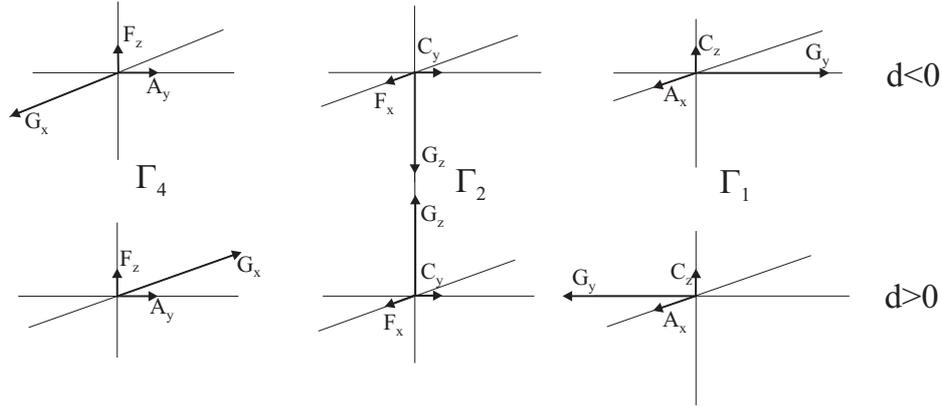}
\caption{Basic vectors of magnetic structure for 3$d$ sublattice in orthoferrites RFeO$_3$ and orthochromites RCrO$_3$}
\label{fig4}
\end{center}
\end{figure*}
 In Table\,\ref{tabler1r2}
we present structural factors $\left[{\bf r}_1\times{\bf r}_2\right]_{x,y,z}$ for the superexchange coupling of the Fe$^{3+}$ ion in position (1/2, 0, 0) with nearest neighbors in orthoferrites with numerical values for \Y . It is easy to see that the weak ferromagnetism in orthoferrites governed by the $y$-component of the Dzyaloshinskii vector does actually make use of only about one-third of its maximal value.

In 1975 we made use of simple formula for the Dzyaloshinskii vector (\ref{d12}) and structural factors from Table\,\ref{tabler1r2} to find a relation between crystallographic and canted magnetic structures for four-sublattice's orthoferrites RFeO$_3$ and orthochromites RCrO$_3$\,\cite{1975,thesis} (see Fig.\,\ref{fig4}), where main G-type antiferromagnetic order  is accompanied by both overt canting characterized by ferromagnetic vector ${\bf F}$ (weak ferromagnetism!) and two types of a hidden canting, ${\bf A}$ and ${\bf C}$ (weak antiferromagnetism!):
$$
F_x=\frac{(x_1+2z_2)ac}{6l^2}\frac{d}{J}G_z\,;\, F_z=-\frac{(x_1+2z_2)ac}{6l^2}\frac{d}{J}G_x\,;\,
$$
$$
A_x=\frac{(\frac{1}{2}+y_2-x_2)ab}{2l^2}\frac{d}{J}G_y;A_y=-\frac{(\frac{1}{2}+y_2-x_2)ab}{2l^2}\frac{d}{J}G_x; $$
\begin{equation}
C_y=\frac{(\frac{1}{2}-y_1)bc}{2l^2}\frac{d}{J}G_z\,;\,C_z=-\frac{(\frac{1}{2}-y_1)bc}{2l^2}\frac{d}{J}G_y\,,
\label{FCA}	
\end{equation}
where $a,b,c$ are unit cell parameters, $x_{1,2}, y_{1,2}, z_2$ are oxygen ($O_{I,II}$) parameters, $l$ is a mean cation-anion separation. These relations imply an averaging on the Fe$^{3+}$\,-\,O$^{2-}$\,-\,Fe$^{3+}$ bonds in $ab$ plane and along $c$-axis. It is worth noting that $|A_{x,y}|>|F_{x,z}|>|C_{y,z}|$.
\begin{table*}
\begin{center}
\caption{Hidden canting in orthoferrites.}
\begin{tabular}{|c|c|c|c|c|c|c|}
\hline
 Orthoferrite  & A$_y$/F$_z$, theory\,\cite{1975} & A$_y$/F$_z$, exp & A$_y$/C$_y$, theory\,\cite{1975} & A$_y$/C$_y$,  exp  \\ \hline
  YFeO$3$ &  1.10 & \begin{tabular}{c}
1.10\,$\pm$\,0.03\cite{Luetgemeier} \\
1.23\,$\pm$\,0.2\cite{Plakhtii}\\
1.1\,$\pm$\,0.1\cite{Georgieva}\\
1.14\,\cite{Park}\\
1.03\,\cite{Amelin}\\
\end{tabular} & 2.04 & ?\\ \hline
  HoFeO$3$ & 1.16
  & 0.85\,$\pm$\,0.10\cite{Georgieva} &  2.00&? \\ \hline
 TmFeO$3$ & 1.10
&  1.25\,$\pm$\,0.05\cite{Luetgemeier}& 1.83 & ?\\ \hline
 YbFeO$3$ & 1.11
&  1.22\,$\pm$\,0.05\cite{Plakhtii} & 1.79 & 2.0\,$\pm$\,0.2\cite{Luetgemeier}\\ \hline
 \end{tabular}
\label{AFC}
\end{center}
\end{table*}

First of all we arrive at a simple relation between crystallographic parameters and magnetic moment of the Fe-sublattice: in units of $G\cdot g/cm^3$
\begin{equation}
	M_{Fe}=\frac{4g\mu_BS}{\rho V}|F_{x,z}|=\frac{2g\mu_BSac}{3l^2\rho V}(x_1+2z_2)\frac{d(\theta)}{J(\theta)}\, ,
\end{equation}
where $\rho$ and $V$ are the unit cell density and volume, respectively.
The overt canting $F_{x,z}$ can be calculated through the ratio of the Dzyaloshinskii ($H_D$) and exchange ($H_E$) fields as follows
\begin{equation}
	F=H_D/2H_E \, .
\end{equation}
If we know the Dzyaloshinskii field we can calculate the $d(\theta )$ parameter in orthoferrites as follows
\begin{equation}
	H_D=\frac{S}{g\mu_B}\sum_i|d_y(1i)|=\frac{S}{g\mu_B}(x_1+2z_2)\frac{ac}{l^2}|d(\theta )|\, ,
\end{equation}
that yields $|d(\theta )|\cong$\,3.2\,K\,=\,0.28\,meV in YFeO$_3$ given $H_D$\,=\,140\,kOe\,\cite{Jacobs}. This  value is in good agreement with the data of recent experiments\,\cite{Park,Amelin} which make it possible to obtain information on the Dzyaloshinskii vectors based on measurements of the spin-wave spectrum.
  It is worth noting that despite $F_z\approx$\,0.01 the $d(\theta )$ parameter is only one order of magnitude  smaller than the exchange integral in YFeO$_3$.

Our results have stimulated experimental studies of the  hidden canting, or "weak antiferromagnetism" in orthoferrites. As shown in Table\,\ref{AFC} the theoretically predicted relations between overt and hidden canting  nicely agree with the experimental data obtained for different orthoferrites by NMR\,\cite{Luetgemeier}, neutron diffraction,  measurement of the low-energy spin excitations by inelastic neutron scattering and by absorption of THz radiation  \,\cite{Plakhtii,Georgieva,Park,Amelin}.

	\subsection{The DM coupling and effective magnetic anisotropy}
	Hereafter we demonstrate a contribution of the DM coupling into effective magnetic anisotropy in orthoferrites.	
	The classical energies of the three spin configurations in orthoferrites $\Gamma_1(A_x,G_y,C_z)$, $\Gamma_2(F_x,C_y,G_z)$, and $\Gamma_4(G_x,A_y,F_z)$ given  $|F_x|=|F_z|=F$, $|C_y|=|C_z|=C$, $|A_x|=|A_z|=A$ can be written as follows\,\cite{thesis}	
	\begin{eqnarray}
E_{\Gamma_1}= E_{G}-48JS^2F^2\left[\frac{1}{3}(\frac{C}{F})^2+\frac{2}{3}(\frac{A}{F})^2\right] \, ; \\
E_{\Gamma_2}= E_{G}-48JS^2F^2\left[1+\frac{1}{3}(\frac{C}{F})^2\right] \, ; \\
E_{\Gamma_4}= E_{G}-48JS^2F^2\left[1+\frac{2}{3}(\frac{A}{F})^2\right] \, ,	
	\end{eqnarray}
	with obvious relation $E_{\Gamma_4} < E_{\Gamma_1}\leq E_{\Gamma_2}$.
	The energies allow us to find the constants of the in-plane magnetic anisotropy $E_{an}=k_1\,cos2\theta$ ($ac$, $bc$ planes), $E_{an}=k_1\,cos2\varphi$ ($ab$ plane): $k_1(ac)=\frac{1}{2}(E_{\Gamma_2}-E_{\Gamma_4})$;  $k_1(bc)=\frac{1}{2}(E_{\Gamma_2}-E_{\Gamma_1})$;
$k_1(ab)=\frac{1}{2}(E_{\Gamma_4}-E_{\Gamma_1})$. Detailed analysis of different mechanisms of the magnetic anisotropy of the orthoferrites\,\cite{thesis,aniso} points to a leading contribution of the DM coupling. Indeed, for all the orthoferrites RFeO$_3$ this mechanism does  predict a minimal energy for $\Gamma_4$ configuration which is actually realized as a ground state for all the orthoferrites, if one neglects the R-Fe interaction. Furthermore, predicted value of the constant  of the magnetic anisotropy in $ac$-plane for YFeO$_3$ $k_1(ac)$\,=2.0$\cdot 10^5$\,erg/cm$^3$ is close enough to experimental value of 2.5$\cdot 10^5$\,erg/cm$^3$\,\cite{Jacobs}.
Interestingly, the model predicts a close energy for $\Gamma_1$ and $\Gamma_2$ configurations so that $|k_1(bc)|$ is about one order of magnitude less than $|k_1(ac)|$ and  $|k_1(ab)|$ for most orthoferrites\,\cite{thesis,aniso}. It means the anisotropy in $bc$-plane will be determined by a competition of the DM coupling with relatively weak contributors such as magneto-dipole interaction and single-ion anisotropy. It should be noted that the sign and value of the $k_1(bc)$ is of a great importance for the determination of the type of the domain walls for orthoferrites in their basic $\Gamma_4$ configuration (see, e.g., Ref.\,\cite{DyFeO3}).

\begin{figure*}[t]
\begin{center}
\includegraphics[width=16cm,angle=0]{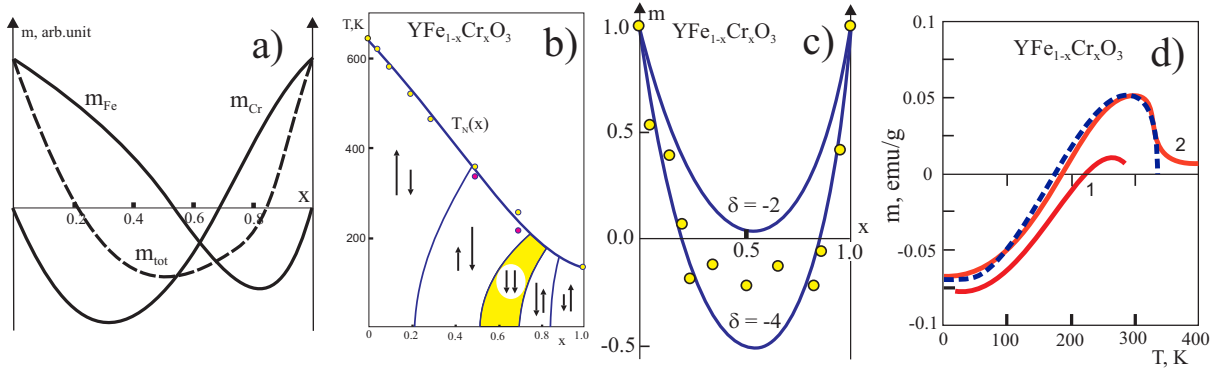}
\caption{(Color online) a) Concentration dependence of the magnetization and Fe-, Cr- partial contributions in YFe$_{0.5}$Cr$_{0.5}$O$_3$; b) The MFA phase diagram of weak ferrimagnet  YFe$_{1-x}$Cr$_x$O$_3$ given $\delta$\,=\,-4; left and right arrows demonstrate the orientation and magnitude of the magnetization for Fe- and Cr-sublattices, respectively.  The outer and inner thin curves mark the compensation points for the net and partial (Fe, Cr) magnetization, respectively. Experimental values of $T_N$ for single crystalline and polycrystalline samples are marked by light  and dark circles, respectively. c) Concentration dependence of the low-temperature magnetization in YFe$_{1-x}$Cr$_x$O$_3$: experimental data (circles)\,\cite{WFIM-1}, the MFA calculations given $\delta$\,=\,-2 and -4; c) ;  d) Temperature dependence of magnetization in YFe$_{1-x}$Cr$_x$O$_3$:  solid curves -- experimental data for x\,=\,0.38 (Kadomtseva {\it et al.}, 1977\,\cite{WFIM-2} -- curve 1) and for x\,=\,0.4 (Dasari {\it et al.}, 2012\,\cite{Dasari} -- curve 2), dotted curve -- the MFA calculation for x\,=\,0.4\,\cite{Dasari} given $d_{FeCr}$\,=\,-0.39\,K }
\label{fig5}
\end{center}
\end{figure*}

\section{Weak ferrimagnetism as a novel type of magnetic ordering in systems with competing signs of the Dzyaloshinskii vector}

First experimental studies of mixed orthoferrites-orthochromites YFe$_{1-x}$Cr$_x$O$_3$\,\cite{WFIM-1} performed in Moscow State University did confirm theoretical predictions regarding the signs of the Dzyaloshinskii vectors and revealed the {\it weak ferrimagnetic} behavior due to a competition of Fe-Fe, Cr-Cr, and Fe-Cr DM coupling with antiparallel orientation of the mean weak ferromagnetic moments of Fe and Cr subsystems in a wide concentration range. In other words, we have predicted a novel class of mixed 3$d$ systems with competing signs of the Dzyaloshinskii vector, so called {\it weak ferrimagnets}. Transversal weak ferromagnetic moment of the Cr$^{3+}$ impurity ion in orthoferrite YFeO$_3$ can be evaluated as follows
\begin{equation}
{\bf m}_{Cr}=g\mu_BS_{Cr}(2\delta -1){\bf F}\, ,
\end{equation}
where dimensionless parameter
\begin{equation}
	\delta =\frac{d_{CrFe}}{d_{FeFe}}\frac{I_{FeFe}}{I_{CrFe}}
\end{equation}
does characterize a relative magnitude of the impurity-matrix DM coupling. By comparing $m_{Cr}$ with that of the matrix value ${\bf m}_{Fe}=g\mu_BS_{Fe}{\bf  F}$ we see that the weak ferromagnetic moment for the Cr  impurity is antiparallel to that of the Fe matrix even for positive but small $\delta <1/2$. However, the effect is more pronounced for negative $\delta$, that is for different signs of $d_{CrFe}$ and $d_{FeFe}$.

\begin{figure}[b]
\begin{center}
\includegraphics[width=8cm,angle=0]{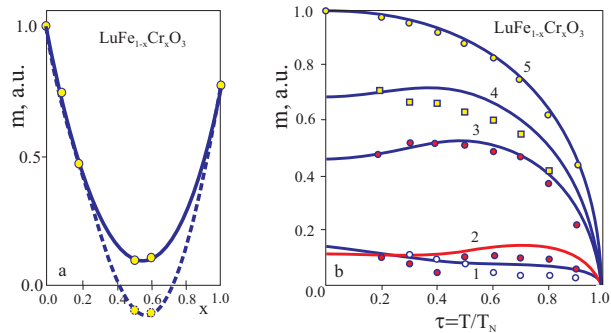}
\caption{ (Color online) a) Concentration dependence of the low-temperature ($T$\,=\,77\,K) magnetization in LuFe$_{1-x}$Cr$_x$O$_3$: experimental data (circles)\,\cite{LuFeCrO3}, the MFA calculations (solid curve) given $\delta$\,=\,-1.5. (b) Temperature dependence of magnetization in LuFe$_{1-x}$Cr$_x$O$_3$: circles -- experimental data\,\cite{LuFeCrO3} given $x$\,=\,0.6 (1), 0.5 (2), 0.2 (3), 0.1 (4), 0.0 (5),  solid curves -- the MFA calculations given $\delta$\,=\,-1.5. }
\label{fig6}
\end{center}
\end{figure}
\begin{figure*}[t]
\begin{center}
\includegraphics[width=12cm,angle=0]{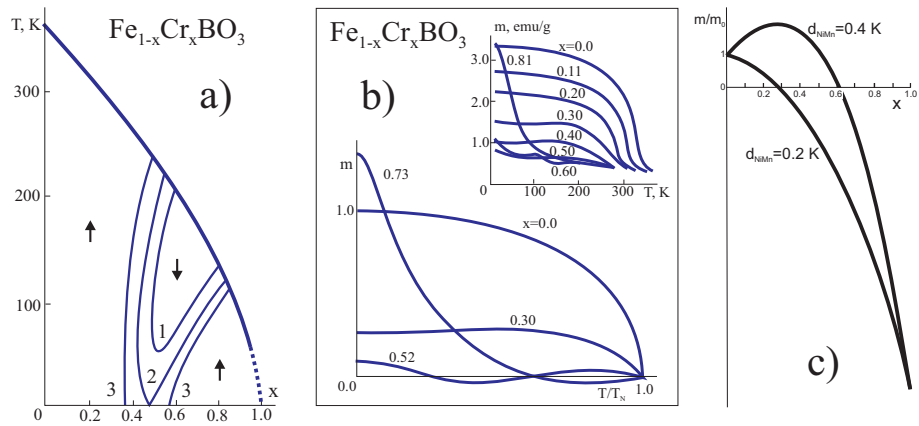}
\caption{a) The MFA simulation of the T-x-phase diagram of the weak ferro(ferri)magnet Fe$_{1-x}$Cr$_x$BO$_3$\,\cite{FeCrBO3} given $I_{FeFe}=I_{FeCr}$=\,-20.3\,K, $I_{CrCr}$\,=\,2.0\,K, arrows point to orientation of the net weak ferromagnetic moment. Curves 1, 2, 3 mark the compensation points given $d_z(FeFe)$\,=\,$d_z(CrCr)$\,=\,0.67\,K, $d_z(FeCr)$\,=\,-0.67\,K (1), -0.75\,K (2), -0.90\,K (3), respectively. b) The MFA simulation of the temperature dependence of the net magnetization in Fe$_{1-x}$Cr$_x$BO$_3$\,\cite{FeCrBO3} given $d_z(FeFe)$\,=\,$d_z(CrCr)$\,=\,-$d_z(FeCr)$\,=\,0.67\,K at different compositions, the insert shows experimental data from Ref.\,\cite{OHoro} taken at external magnetic field 1\,T. c) The MFA simulation of the concentration dependence of the low-temperature magnetization in Mn$_{1-x}$Ni$_x$CO$_3$\,\cite{MnNiCO3} given $d_z(MnNi)>d_z^{(0)}(MnNi)$ and $d_z(MnNi)<d_z^{(0)}(MnNi)$, respectively.}
\label{fig7}
\end{center}
\end{figure*}

Results of a simple mean-field calculation presented  in Figs.\ref{fig5}-\ref{fig7} show that the weak ferrimagnets such as RFe$_{1-x}$Cr$_x$O$_3$, Mn$_{1-x}$Ni$_x$CO$_3$, Fe$_{1-x}$Cr$_x$BO$_3$\,\cite{WFIM-1,WFIM-2,LuFeCrO3,FeCrBO3,FeNi,MnNiCO3} do reveal very nontrivial concentration and temperature dependencies of magnetization, in particular, the compensation point(s).

In Fig.\,\ref{fig5}a,b,c we do present the MFA "weak ferrimagnetic"\, phase diagram,  concentration dependence of the low-temperature net magnetization, and the Fe, Cr partial contributions  in YFe$_{1-x}$Cr$_x$O$_3$ calculated at constant value of  $\delta$\,=\,-4. Comparison with experimental data for the low-temperature net magnetization\,\cite{WFIM-1} and the MFA calculations with $\delta$\,=\,-2 (Fig.\,\ref{fig5}c) points to a reasonably well agreement everywhere except $x\sim$\,0.5, where $\delta$ parameter seems to be closer to $\delta$\,=\,-3.
In Fig.\,\ref{fig5}d we compare first pioneering experimental data for the temperature dependence of magnetization $m(T)$ in weak ferrimagnet YFe$_{1-x}$Cr$_x$O$_3$ ($x$\,=\,0.38) (Kadomtseva {\it et al.}, 1978\,\cite{WFIM-2} -- curve 1) with recent  data for a close composition with $x$\,=\,0.4 (Dasari {\it et al.}, 2012\,\cite{Dasari} -- curve 2). It is worth noting that recent MFA calculations by Dasari {\it et al}.\,\cite{Dasari}  given $d_{FeCr}$\,=\,-0.39\,K provide very nice description of $m(T)$ at $x$\,=\,0.4. Note that the authors\,\cite{Dasari} found a rather strong dependence of the $d_{FeCr}$ parameter on the concentration $x$.
The concentration and temperature dependencies of magnetization in LuFe$_{1-x}$Cr$_x$O$_3$ are nicely described by a simple MFA model under the assumption of constant sign magnetization given constant value of $\delta$\,=\,-1.5 (Fig.\,\ref{fig6}a,b\,\cite{LuFeCrO3}),
  which, strictly speaking, did not exclude the possibility of an alternative description of the dependence $m(x)$ with two points of concentration compensation of magnetization (see dotted line in Fig.\,\ref{fig6}a). Furthermore, strictly speaking, the absence of compensation concentration points for low-temperature magnetization $m(x, T = 77 K)$ does not mean the absence of compensation points at higher temperatures.  Indeed,  much later, in 2016, Pomiro {\it et al}.\,\cite{Pomiro} observed the spontaneous magnetization reversal in polycrystalline LuFe$_{0.5}$Cr$_{0.5}$O$_3$ below T$_N$\,=\,290\,K at a compensation temperature T$_{comp}$\,=\,224\,K and
  Billoni {\it et al.}\,\cite{Billoni} have performed more advanced classical Monte Carlo simulations  for
RFe$_{1-x}$Cr$_x$O$_3$ with R = Y and Lu, comparing the numerical simulations with experiments
and MFA calculations. In addition to the dependence T$_N(x)$, this model is able to reproduce  the magnetization reversal (MR) observed experimentally in a field cooling process for intermediate $x$ values.
At variance with YFeO$_3$ and YCrO$_3$ which are weak ferromagnets with main G$_x$F$_z$-type magnetic structure below T$_N$, the orthoferrites-orthochromites YFe$_{1-x}$Cr$_x$O$_3$, which
referred as weak ferrimagnets, reveal full or partial G$_x$F$_z$--\,G$_z$F$_x$ type spin-reorientation in a wide range of substitution.
This unexpected behavior which is usually typical for orthoferrites with magnetic rare-earth ions (Er, Tm, Dy,…) was attributed mainly to a strong decrease of the DM contribution to magnetic anisotropy in the $ac$-plane for $x$\,=\,0.5--0.6\,\cite{Agafonov,RFeCrO3-2}. In contrast to the yttrium system, the lutetium orthoferrite-orthocromites LuFe$_{1-x}$Cr$_x$O$_3$ (x = 0.0, 0.1, 0.2, 0.5, 0.6, and 1.0) reveal the main G$_x$F$_z$ type magnetic structure without signatures of the spontaneous spin-reorientation transition.
This difference can be explained by the significantly larger contribution of single-ion anisotropy to $k_{ac}$ in LuFeO$_3$ as compared to YFeO$_3$\,\cite{thesis,Mukhin}.

Let us turn to the features of other weak ferrimagnets.
Fig.\,\ref{fig7}b shows a calculated phase diagram of the trigonal weak ferrimagnet Fe$_{1-x}$Cr$_x$BO$_3$\,\cite{FeCrBO3}.
Temperature-dependent magnetization studies from 4.2 to 600\,K have been made for the solid solution system Fe$_{1-x}$Cr$_x$BO$_3$ where $0\leq x \leq $\,95\,\cite{OHoro}. A rapid decrease is observed in the saturation magnetization with increasing $x$ at 4.2\,K up to 0.40, after which a broad compositional minimum is found up to x=0.60. Compositions in the range of 0.40$\leq x \leq$0.60 display unusual magnetization behavior as a function of temperature in that maxima and minima are present in the curves below the Curie temperatures. Fig.\,\ref{fig7}b shows a nice agreement between experimental data\,\cite{OHoro} and our MFA calculations.

At variance with the $d^5$\,-\,$d^3$ (Fe-Cr) mixed systems such as YFe$_{1-x}$Cr$_x$O$_3$ or Fe$_{1-x}$Cr$_x$BO$_3$ the manifestation of different DM couplings Fe-Fe, Cr-Cr, and Fe-Cr in (Fe$_{1-x}$Cr$_x$)$_2$O$_3$  is all the more surprising because of different magnetic structures of the end compositions, $\alpha$-Fe$_{2}$O$_3$ and Cr$_{2}$O$_3$ and emergence of a nonzero DM coupling for the three-corner-shared FeO$_6$ and CrO$_6$ octahedra, "forbidden"\, for Fe-Fe and Cr-Cr bonding. All this makes magnetic properties of mixed compositions (Fe$_{1-x}$Cr$_x$)$_2$O$_3$ to be very unusual\,\cite{FeCrO}.


Unlike the $d^5$\,-\,$d^3$ (Fe-Cr) mixed systems  YFe$_{1-x}$Cr$_x$O$_3$ or Fe$_{1-x}$Cr$_x$BO$_3$, where the two concentration compensation points do occur given rather large $d_{FeCr}$ parameter, in the $d^5$\,-\,$d^8$   systems, the nickel and fluorine substituted orthoferrites RFe$_{1-x}$Ni$_x$F$_y$O$_{3-y}$\,\cite{FeNi} or  Mn$_{1-x}$Ni$_x$CO$_3$ with Mn$^{2+}$\,-\,Ni$^{2+}$ pairs\,\cite{MnNiCO3} we have the only concentration compensation point irrespective of the  $d_{MnNi}$ parameter. However, the character of the concentration dependence of the weak ferrimagnetic moment $m(x)$ depends strongly on its magnitude.
Given the increasing concentration the $m(x)$  is first rising or falling  with $x$ depending on whether $d_{MnNi}$ greater than, or less than $d_{MnNi}^{(0)}=(1+\frac{S_{Mn}}{S_{Ni}})\frac{I_{MnNi}}{2I_{MnMn}}d_{MnMn}$. Figure (\ref{fig7}c) does clearly demonstrate this feature.

It should be noted that just recently Dmitrienko {\it et al.}\,\cite{Dmitrienko-EASTMAG} have first discovered experimentally that in accordance with our theory (see Table\,\ref{tablesign}) the sign of the Dzyaloshinskii vector in MnCO$_3$ ($d^5$\,-\,$d^5$) coincides with that one in FeBO$_3$ ($d^5$\,-\,$d^5$) , whereas NiCO$_3$ ($d^8$\,-\,$d^8$)   demonstrates the opposite sign.

\subsection{Features of the 4$f$\,-\,3$d$ interaction in weak ferrimagnets RFe$_{1-x}$Cr$_x$O$_3$}

It is undoubtedly of interest to investigate the influence
of the weak ferrimagnetic ordering of the 3$d$-sublattice on the
behavior of the rare-earth subsystem  in mixed ferrites-chromites
RFe$_{1-x}$Cr$_x$O$_3$. The character of polarization of the R-ions and its concentration and temperature dependencies yield valuable information not only on the state of the $d$-subsystem, but also on the 4$f$\,-\,3$d$ interaction mechanisms, primarily on the relative roles of the ferro- and antiferromagnetic contributions to the effective field at the R-ions\,\cite{GdFeCrO3,RFeCrO3-1,RFeCrO3-2,RFeCrO3-3}.
 Of particular interest, in our opinion, is the
GdFe$_{1-x}$Cr$_x$O$_3$ system with $S$-type 4$f$- and 3$d$-ions, where it
might seem that it is precisely the ferromagnetic contribution due to the isotropic 4$f$\,-\,3$d$ exchange should play the decisive role in the polarization of the Gd sublattice. At the same time, a detailed analysis of the magnetic properties of
GdFeO$_3$ and GdCrO$_3$\,\cite{Belov,GdFeCrO3}  has quite unexpectedly revealed the substantial role of the anisotropic exchange of the $S$-ions Gd$^{3+}$ with $S$-type ions  Fe$^{3+}$ and Cr$^{3+}$ and accordingly of the antiferromagnetic contribution to the polarization of the Gd sublattices with the predominance of antisymmetric term which is determined by the 4$f$\,-\,3$d$ DM coupling (see (\ref{Fe}) and (\ref{Cr})).

\begin{figure}[t]
\begin{center}
\includegraphics[width=8.5cm,angle=0]{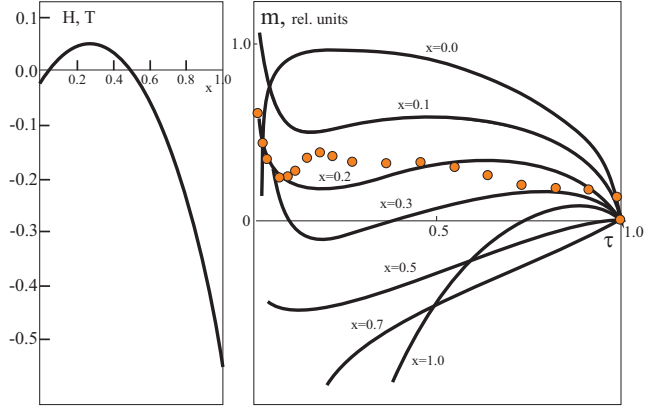}
\caption{(Color online) Left panel: Calculated concentration dependence of the average effective field on the Gd$^{3+}$ ions in the weak ferrimagnet  at T\,=\,0\,K. Right panel: Calculated temperature dependencies of the spontaneous magnetization for
a number of weak ferrimagnets GdFe$_{1-x}$Cr$_x$O$_3$ and end compositions, $\tau$\,=\,T/T$_N$. Solid circles are experimental data for single crystals GdFe$_{0.83}$Cr$_{0.17}$O$_3$ with T$_N$\,=\,550\,K\,\cite{GdFeCrO3}.}
\label{fig8}
\end{center}
\end{figure}

Knowledge of the numerical values of the parameters
of isotropic and anisotropic 4$f$\,-\,3$d$ interaction has enabled us to calculate, within the
framework of the molecular-field theory, the concentration
and temperature dependencies of the average effective field
$H_z$, the magnetization of the Gd sublattice, and the total
magnetization of GdFe$_{1-x}$Cr$_x$O$_3$ in the entire range of concentrations
$x$ (see Fig.\ref{fig8})\,\cite{GdFeCrO3}. The exchange integrals and the DM coupling
parameters in the $d$ sublattices were chosen equal
to the corresponding values for weak ferrimagnet YFe$_{1-x}$Cr$_x$O$_3$.
The concentration dependence $H_z(x)$
 at T\,=\,0\,K  has very unusual form with two
compensation points, at small and relatively large concentrations
of the Cr$^{3+}$ ions. Whereas at $x\approx$\,0.05 the compensation
of the total magnetic moment is still observed, at
$x$\,=\,0.10 the reversal of the sign of $H_z$ leads to a hyperbolic
increase of $m(T)$ in the low-temperature region. At the same time the
calculation shows that at $x\approx$\,0.27 and $\tau\approx$\,0.17 we arrive at the compensation point, which then "bifurcates" with one (high-temperature) compensation point moving
towards T$_N$, with increasing $x$, while the other (low temperature)
towards T\,=\,0\,K. At $x$\,>\,0.5 the compensation
points vanish. Only for compositions directly adjacent to
pure gadolinium orthochromite is the compensation again
observed, and with increasing concentration of the Fe$^{3+}$
ions the compensation point shifts from T$_{comp}$\,=\,110\,K in pure
GdCrO$_3$, to T$_{comp}$\,=T$_N$, at $x\approx$\,0.95.
As a whole the calculated concentration and temperature dependencies of the magnetization
in GdFe$_{1-x}$Cr$_x$O$_3$ agree satisfactorily
with the experimental data\,\cite{GdFeCrO3}.
Finally, we note the need for further experimental investigation
of the rare-earth weak ferrimagnets such as GdFe$_{1-x}$Cr$_x$O$_3$
 both from the viewpoint of studying various
$f$\,-\,$d$ interactions, and of the possibility of obtaining novel advanced
magnetic properties.

\subsection{Unconventional spin-reorientation in weak ferrimagnets }

\begin{figure}[t]
\begin{center}
\includegraphics[width=7.5cm,angle=0]{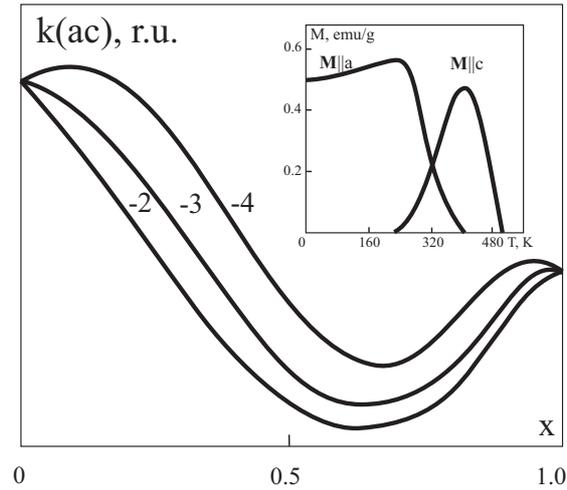}
\caption{Concentration dependence of the DM coupling contribution to the first anisotropy constant in $ac$-plane given different values of the parameter $\delta$. Inset: Temperature dependence of the magnetization in the weak ferrimagnet  YFe$_{0.85}$Cr$_{0.15}$O$_3$\,\cite{RFeCrO3-2} demonstrating the $\Gamma_4$\,--\,$\Gamma_2$ spin-reorientation transition in the temperature range 240\,-\,400\,K.}
\label{fig9}
\end{center}
\end{figure}

The contribution of the competing antisymmetric exchange to the magnetic anisotropy of weak ferrimagnets has an unusual concentration dependence. So, if in pure orthoferrite YFeO$_3$ and orthochromite YCrO$_3$ antisymmetric exchange makes a decisive contribution to the stabilization of the magnetic configuration $\Gamma_4$, then in a weak ferrimagnet YFe$_{1-x}$Cr$_x$O$_3$ it can induce a spin-reorientation $\Gamma_4$\,--\,$\Gamma_2$ transition which are typical for several orthoferrites RFeO$_3$ with magnetic rare-earth ions (R = Nd, Sm, Tb, Ho, Er, Tm, Yb). In Fig.\,\ref{fig9} we demonstrate concentration dependence of the DM coupling contribution to first anisotropy constant for YFe$_{1-x}$Cr$_x$O$_3$ in $ac$-plane given different values of the parameter $\delta$, which was calculated within simple mean-field approximation\,\cite{Agafonov,RFeCrO3-2} in the limit of low temperatures. A characteristic feature of this dependence is the appearance of several extrema with a sharp decrease in the contribution in the region of intermediate concentrations near $x$\,$\sim$\,0.6\,-\,0.7. Furthermore, similarly magnetization, this contribution to anisotropy  has a specific temperature dependence\,\cite{Agafonov,RFeCrO3-2}. On the whole, both effects can lead to the appearance of spontaneous spin-reorientation transitions in weak ferrimagnets of the YFe$_{1-x}$Cr$_x$O$_3$  type with a nonmagnetic "R"-ion. Indeed, in full accordance with the theory, such transitions have been observed experimentally, for example, the $\Gamma_4$\,--\,$\Gamma_2$ spin-reorientation transition in YFe$_{0.85}$Cr$_{0.15}$O$_3$\,\cite{Agafonov,RFeCrO3-2} (see Fig. \,\ref{fig9}).

A more surprising situation was found in the weak ferrimagnet DyFe$_{1-x}$Cr$_x$O$_3$ at a relatively low concentration of Cr ions. The Dy$^{3+}$ ions in DyFeO$_3$ stabilize the $\Gamma_1(G_y)$ configuration, so that at T\,=\,40\,K, a jump-like Morin transition $\Gamma_4$\,--\,$\Gamma_1$ is observed. In all single crystals of weak ferrimagnets  DyFe$_{1-x}$Cr$_x$O$_3$ ($x$\,=\,0.07, 0.10, 0.13, 0.15, 0.36, 0.40) synthesized and studied in the laboratory of A.~M. Kadomtseva (Moscow State University)\,\cite{Agafonov,RFeCrO3-2}, the Morin $\Gamma_4$\,--\,$\Gamma_1$ spin-reorientation transition to the low-temperature phase $\Gamma_1$ was detected, with the exception of the composition with $x$\,=\,0.36, where the phase $\Gamma_2$ was unexpectedly found to be the high-temperature phase. Puzzlingly, in compositions $x$\,=\,0.1 and $x$\,=\,0.13, the Morin transition proceeded according to the $\Gamma_4$\,--\,$\Gamma_{421}$\,--\,$\Gamma_{21}$\,--\,$\Gamma_1$ ($x$\,=\,0.1) or $\Gamma_4$\,--\,$\Gamma_{421}$\,--\,$\Gamma_1$ ($x$\,=\,0.1) scheme and was accompanied by the deviation of the antiferromagnetic vector ${\bf G}$ into space with the appearance in a narrow temperature range of the projection of the magnetic moment on the $b$-axis. Never before has the state of the mixed configuration $\Gamma_{421}(G_xG_yG_z)$ with the spatial orientation of the antiferromagnetism vector and the appearance of the $b$-component of the magnetic moment ($M_b\propto G_xG_yG_z$) been observed.

\subsection{Recent renewal of interest to weak ferrimagnets}
 The systems with competing antisymmetric exchange were extensively investigated up to the end of 80ths mainly in the laboratory of A.~M. Kadomtseva at Moscow State University. Recent renewal of interest to the systems with the compensation point was stimulated by  the perspectives of the applications in magnetic memory (see, e.g., Refs.\,\cite{Mao,Dasari} and references therein).
For instance, weak ferrimagnet YFe$_{0.5}$Cr$_{0.5}$O$_3$ exhibits magnetization reversal at low applied fields. Below a compensation temperature (T$_{comp}$), a tunable bipolar switching of magnetization is demonstrated by changing the magnitude of the field while keeping it in the same
direction. The compound also displays both normal and inverse magnetocaloric effects above and below 260\,K, respectively. These phenomena coexisting in a single magnetic system can be tuned in a predictable manner and have potential applications in electromagnetic devices\,\cite{Mao}.
Weak ferrimagnets can exhibit the tunable exchange bias (EB) effect\,\cite{Bora}.
Recently the EB effect with reversal sign was found in LuFe$_{0.5}$Cr$_{0.5}$O$_3$ ferrite-chromite\,\cite{EB} which is a weak ferrimagnet below T$_N$\,=\,265\,K, exhibiting antiparallel orientation of the mean weak ferromagnetic moments of the Fe and Cr sublattices due to opposite sign of the Fe\,-\,Cr Dzyaloshinskii vector as compared to that of the Fe\,-\,Fe and Cr\,-\,Cr. The weak FM moments of the studied compound compensate each other at temperature T$_{comp}$\,=\,230\,K, leading to the net magnetic moment reversal and to observed negative magnetization, at moderate applied field, below T$_{comp}$. Variety of such extraordinary properties as high compensation temperature, temperature-controlled positive/negative EB below/above T$_{comp}$, and switching the magnetization direction to the opposite one by magnetic field without of changing its polarity makes weak ferrimagnet LuFe$_{0.5}$Cr$_{0.5}$O$_3$  of promising candidate for application in magnetic memories.

Combining magnetization reversal effect with magnetoelectronics can exploit tremendous technological potential for device applications, for example, thermally assisted magnetic random access memories, thermomagnetic switches and other multifunctional devices, in a preselected and convenient manner.
Nowadays a large body of magnetic materials might be addressed as systems with competing antisymmetric exchange\,\cite{Kumar}, including novel class of mixed helimagnetic B20 alloys such as Mn$_{1-x}$Fe$_x$Ge where the helical nature of the main ferromagnetic spin structure is determined by a competition of the DM couplings Mn\,-\,Mn, Fe\,-\,Fe, and Mn\,-\,Fe. Interestingly, that the magnetic chirality in the mixed compound changes its sign at $x_{cr}\approx$\,0.75, probably due to different sign of the Dzyaloshinskii vectors for Mn\,-\,Mn and Fe\,-\,Fe pairs\,\cite{MnFeGe}.

\section{Determination of the sign of the Dzyaloshinskii vector}

Determining the sign of the Dzyaloshinskii vector and the relative orientation of the ${\bf F}$ and ${\bf G}$ vectors in weak ferromagnets is of both fundamental importance from the standpoint of the microscopic theory of the DM coupling and practical importance for the reliable identification of the parameters of various anisotropic interactions in these materials. In particular, for the rare-earth orthoferrites RFeO$_3$, this concerns the parameters of the 4$f$\,-\,3$d$ interaction\,\cite{Belov}, the parameters of transferred  and supertransferred hyperfine interactions\,\cite{Moskvin-ASTHF}, and the magnitude of the effective magnetic field for $\mu$-mesons\,\cite{muon}.
The sign of the Dzyaloshinskii vector determines the handedness of spin helix in crystals with the noncentrosymmetric B20 structure.

How to measure the sign of the DM interaction in weak ferromagnets? According to Ref.\,\cite{Ozhogin}, an answer to this question can be given by determining experimentally the direction of rotation of the antiferromagnetism vector ${\bf l}$ around the magnetic field ${\bf H}$  in the geometry ${\bf H}\parallel {\bf d}\parallel$easy axis. However, as was pointed out later (see Ref.\,\cite{Chep}), a M\"{o}ssbauer experiment  on easy-axis hematite did not give an unambiguous result.

According to Dmitrienko {\it et al.}\,\cite{Dmitrienko}, first of all, a strong enough magnetic field should be applied to obtain the single domain state where the DM coupling pins antiferromagnetic ordering to the crystal lattice. Next, single crystal diffraction methods sensitive both to oxygen coordinates and to the phase of antiferromagnetic ordering should be used. In other words, one should observe those Bragg reflections $hkl$ where interference between magnetic scattering on Mn atoms and nonmagnetic scattering on oxygen atoms is significant. There are three suitable techniques: neutron diffraction, M\"{o}ssbauer $\gamma$-ray diffraction, and resonant x-ray scattering. The sign of the DM vector in weak ferromagnetic FeBO$_3$ was deduced from observed interference between resonant X-ray scattering and magnetic X-ray scattering\,\cite{Dmitrienko}.

The authors in Ref.\,\cite{Chep} claimed that the character of the field-induced transition from an antiferromagnetic phase to a canted phase in  cobalt fluoride CoF$_2$ is due to the "sign" of the Dzyaloshinskii interaction, and this affords an opportunity to determine experimentally the sign of the Dzyaloshinskii vector. However, in fact they addressed a symmetric Dzyaloshinskii interaction that is magnetic anisotropy
$$
V_{sym}=-D(m_xl_y+m_yl_x)
$$
 rather than antisymmetric DM coupling.

In our opinion, the most reliable experimental method for determining the mutual orientation of the vectors of ferromagnetism ${\bf F}$ and antiferromagnetism ${\bf G}$, and hence the sign of the Dzyaloshinskii vector, is to study the magnitude and sign of the effective magnetic field on ligands, as well as $\mu$-mesons in weak ferromagnets.

 \subsection{Positive muons in
orthoferrites as a tool to examine the sign of Dzyaloshinskii vector}

In muon spin rotation ($\mu$SR) experiments, spin-polarized
positive (anti)muons are used to probe the microscopic field
distribution at the interstitial site(s) where the $\mu^+$ stop inside
the sample under investigation. The extreme sensitivity of
the muon to small magnetic fields as well as the absence of
quadrupolar coupling makes this technique very promising in
probing magnetic orders, offering a valuable alternative to
neutron scattering. This approach, which shares many similarities with nuclear magnetic resonance, has the advantage
of being applicable to virtually any material, but it has the
drawback that the interstitial sites where the muon stops and
the nature of muon interaction with the host are generally
unknown\,\cite{mu1}.
Site assignment is the key initial ingredient in the not
infrequent cases where the internal magnetic field is dominated
by the distant dipole contribution, which requires only the
knowledge of the site in order to be computed by a classical
sum over the dipole moments of the host lattice.
Thus, the comparison between predicted and measured local
field can validate the muon site assignment, and in turn, this
assessment yields, e.g., a measure of the magnetic moment
values.  However, additional local field contributions exist and
these are not negligible in many cases\,\cite{mu2}.

The dipolar field can be approximated with good accuracy, assuming a classical moment ${\bf M}$ centered at the atomic positions of the magnetic atoms, and evaluating the total contribution as
\begin{equation}
{\bf H}({\bf r}_{\mu})=\sum_j\left(\frac{3{\bf r}_{\mu j}({\bf M}_j\cdot{\bf r}_{\mu j})}{r_{\mu j}^5}-\frac{{\bf M}_j}{r_{\mu j}^3}\right)\, ,
\label{dip}
\end{equation}
where
${\bf r}_{\mu}$ is  the muon position, ${\bf M}_j$ is the magnetic moment of $j$-th  ion, ${r_{\mu j}}$ is the distance between $j$-th ion and the muon site.
The hyperfine Fermi contact field contribution, transferred or supertransferred, can be written as follows
\begin{equation}
 H({\bf r}_{\mu})=\frac{8\pi}{3}\mu_B\rho_s({\bf r}_{\mu})\, ,
\label{cont}
\end{equation}
where  $\rho_s$ is the spin density at the muon site\,\cite{mu2}.

 First detailed investigation of positive muons in orthoferrites RFeO$_3$ (R= Sm, Eu, Dy, Ho, Y, Er) has been performed by Holzschuh {\it et al.}\,\cite{muon}.
 In their presentation the hyperfine
field at the muon site in the orthoferrites can be explained
in terms of dipolar fields  only.
By comparing measured internal magnetic fields with calculated dipolar fields of  Fe$^{3+}$ ions the authors found the position of stable muon site, furthermore, they  established that in configuration $\Gamma_4$ the sign of $G_x$ should be positive   for $F_z>0$, in accordance with our earlier theoretical predictions\,\cite{1977}, since only this assumption leads to a reasonable muon site.

However, these results were severely criticized in work\,\cite{muon-1}, the authors of which argued that the interpretation\,\cite{muon}
contains some serious flaws: important details have not
been worked out correctly and their analysis is not complete
enough to support some of their conclusions. First of all it concerns the supertransferred hyperfine field contribution, which must not be disregarded. Furthermore,  they drew attention to the need for strict accounting of the sign convention, labeling of the Fe$^{3+}$ ions and the representation of the spin configurations which is not uniform in the literature. This all casts doubt in using theoretical relations, in particular, concerning
the mutual orientation of the ferro- and antiferromagnetic vectors, that is, in fact, the sign of the Dzyaloshinskii vector.

\subsection{The ligand NMR in weak ferromagnets and first reliable determination of the sign of the Dzyaloshinskii vector}

As was firstly shown in our paper\,\cite{sign} reliable local information on the sign of the Dzyaloshinskii vector, or to be exact, that of the Dzyaloshinskii parameter $d_{12}$, can be extracted from the ligand NMR data in weak ferromagnets. The procedure was described in details for $^{19}$F NMR data in weak ferromagnet FeF$_3$\,\cite{sign}.

\begin{figure}[t]
\centering
\includegraphics[width=8.5cm,angle=0]{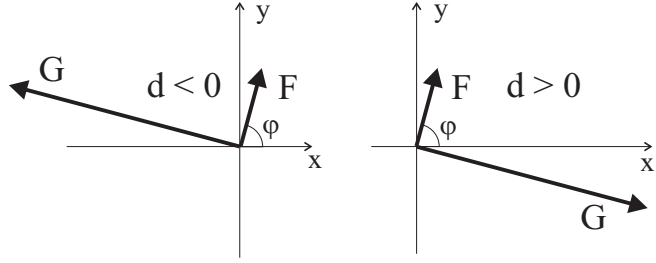}
\caption{Two mutual orientations of $\bf F$ and $\bf G$ vectors in basal plane of FeF$_3$.}
\label{fig10}
\end{figure}

The F$^-$ ions in the unit cell of FeF$_3$ occupy the six positions\,\cite{Hepworth}.
In a trigonal basis these are $\pm(x,1/2-x,1/4),\quad\pm(1/2-x,1/4,x),
\pm(1/4,x,1/2-x)$, that correspond to i) $\pm(3p(x-1/4),\sqrt{3}p
(1/4-x),c/4)$, ii) $\pm(3p(1/4-x),\sqrt{3}p(1/4-x),c/4)$,
and iii) $\pm(0,2\sqrt{3}p(x-1/4),c/4)$ in an orthogonal basis with $O_z\parallel C_3$ and $O_x\parallel C_2$. Each F$^-$ ion is surrounded by two Fe$^{3+}$ from different magnetic sublattices. Hereafter we introduce basic ferromagnetic {\bf F} and antiferromagnetic {\bf G} vectors:
\begin{equation}
\label{m6}
2S{\bf F}={\bf S_1}+{\bf S_2}, 2S{\bf G}={\bf S_1}-{\bf S_2},
{\bf F^2}+{\bf G^2}=1,
\end{equation}
where Fe$_1^{3+}$ and  Fe$^{3+}_2$ occupy positions (1/2,1/2,1/2) and (0,0,0), respectively. FeF$_3$ is an easy plane weak ferromagnet with {\bf F} and {\bf G} lying in (111) plane  with ${\bf F}\bot{\bf G}$.
The two possible variants of the mutual orientation of the {\bf F} and {\bf G}
vectors in the basis plane,   tentatively called as "left" and "right", respectively,  are shown in Fig.\,\ref{fig10}.
The DM energy per Fe$^{3+}$\,-\,F$^-$\,-\,Fe$^{3+}$ bond can be written as follows
$$
	E_{DM}=-2S^2d_z(12)(F_xG_y-F_yG_x)=
$$
\begin{equation}
-\frac{4\sqrt{3}}{l^2}p^2(x+\frac{1}{4})d(\theta )=+0.78S^2d(\theta )(F_xG_y-F_yG_x) \, .
\end{equation}
In other words, the "left" and "right" orientations of basic vectors are realized at $d(\theta )<0$ and $d(\theta )>0$, respectively.

Absolute magnitude of the ferromagnetic vector is numerically equals to an overt canting angle which can be found making use of familiar values of the Dzyaloshinskii field: $H_D=48.8$\,kOe and exchange field: $H_E=4.4\cdot
10^3$\,kOe\,\cite{Prozorova} as follows
\begin{equation}
F=H_D/2H_E\simeq 5.5\cdot 10^{-3}.
\end{equation}
If we know the Dzyaloshinskii field we can calculate the $d(\theta )$ parameter as follows
\begin{equation}
	H_D=\frac{6S}{g\mu_B}|d_z(12)|=\frac{6S}{g\mu_B}0.39|d(\theta )|=48.8\,kOe \, ,
\end{equation}
that yields $|d(\theta )|\cong$\,1.1\,K that is three times smaller than in YFeO$_3$.

The local field on the nucleus of the nonmagnetic F$^-$ anion in weak ferromagnet FeF$_3$, induced by neighboring magnetic $S$-type ion (Fe$^{3+}$, Mn$^{3+}$,\ldots) can be written as follows\,\cite{Turov}
\begin{equation}
\label{m1}
{\bf H}=-\frac{1}{\gamma_n}\hat A{\bf S}
\end{equation}
($\gamma_n$ is a gyromagnetic ratio, $\gamma_n$=4.011 MHz/kOe, ${\bf S}$ is the spin moment of the magnetic ion), where the tensor of the transferred hyperfine interactions (HFI)
$\hat A$ consists of two terms: i) an isotropic contact term with $A_{ij}=A_s\delta_{ij}$
\begin{equation}
\label{m2}
A_s=\frac{f_s}{2S}A^{(0)}_s,\qquad A^{(0)}_s=\frac{16}{3}\pi\mu_B\gamma_n
|\varphi_{2s}(0)|^2\,  ;
\end{equation}
ii) anisotropic term with
\begin{equation}
A_{ij}=A_p(3n_in_j-\delta_{ij}),
\end{equation}
where ${\bf n}$ is a unit vector along the nucleus-magnetic ion bond and the
 $A_p$ parameter includes the dipole and covalent contributions
\begin{equation}
\label{m4}
A_p=A_p^{cov}+A_d,
\end{equation}
$$
A_p^{cov}=\frac{(f_\sigma-f_\pi)}{2S}A_p^{(0)},
A_p^{(0)}=\frac{4}{5}\mu_B\gamma_n\langle\frac{1}{r^3}\rangle_{2p},
$$
\begin{equation}
\label{m5}
A_d=\frac{g_s\mu_B\gamma_n}{R^3}.
\end{equation}
Here $f_{s,\pi,\sigma}$ are parameters for the spin density transfer: magnetic ion - ligand along the proper $s-, \sigma-, \pi$-bond\,\cite{Jacobson};
$|\varphi_{2s}(0)|^2$ is a probability density of the $2s$-electron on nucleus; $\langle \frac{1}{r^3}\rangle_{2p}$ is a radial average.

The transferred HFI for $^{19}$F in fluorides are extensively studied by different techniques, NMR, ESR, and ENDOR\,\cite{Turov}. For $^{19}$F one observes large values both of $A_s^{(0)}$ and $A_p^{(0)}$; $A_s^{(0)}=4.54\cdot~10^4$,
$A_p^{(0)}=1.28\cdot~10^3$ MHz\,\cite{Turov}, together with the 100\% abundance, nuclear spin $I=1/2$, and large   gyromagnetic ratio
this makes the study of the transferred HFI to be simple and available one.

Contribution of the isotropic and anisotropic transferred HFI to the local field on the $^{19}$F can be written as follows
$$
{\bf H}(iso)=-\frac{2S}{\gamma_n}A_s{\bf F}=a_F{\bf F},
$$
\begin{equation}
{\bf H}(an)={\bf \stackrel{\leftrightarrow}{a}}{\bf G}, \hat a=-\frac{2S}{\gamma_n}({\bf \stackrel{\leftrightarrow}{A}}(1)-{\bf \stackrel{\leftrightarrow}{A}}(2)).
\end{equation}
The $A_s$ and $A_p$ parameters we need to calculate parameter $a_F$ and the HFI anisotropy tensor ${\bf \stackrel{\leftrightarrow}{a}}$ that is to calculate the "ferro-" and "antiferro-" contributions to $H$ one can find in the literature data for the pair $^{19}$F\,-\,Fe$^{3+}$. For instance, in KMgF$_3$:Fe$^{3+}$ ($R_{MgF}$\,=\,1.987\AA)\,\cite{Hall} $A_s=+72, A_p=+18$\,MHz, in K$_2$NaFeF$_6$ ($R_{FeF}$\,=\,1.91\AA),
 in K$_2$NaAlF$_6$:Fe$^{3+}$ $A_s$\,=\,+70.17, $A_p$\,=\,+20.34\,MHz\,\cite{Adam}. Thus, we expect in FeF$_3$ $|a_F|\sim 350\div
360$\,MHz $(a_F<0)$ and $H(iso)\simeq 2$\,MHz ($\simeq 0.5$\,kOe).


In the absence of an external magnetic field the NMR frequencies for $^{19}$F in positions 1,2,3 can be written as follows
$$
\nu^2=\gamma_n^2[(\hat a{\bf G})^2+(a_f{\bf F})^2+2a_F{\bf F}\hat a{\bf G}]=
$$
$$
\gamma_n^2(a_{xy}^2+a_F^2F^2\pm 2a_Fa_{xy}F)+
$$
\begin{equation}
\gamma_n^2(a_{yz}^2\mp 4a_Fa_{xy}F)\left\{
\parbox{3cm}{$\cos^2\varphi \\
\cos^2(\varphi+60^o)\\ \cos^2(\varphi-60^o)$}\right.
\label{m9}
\end{equation}
where the $a_{xy},a_{yz}$ components are taken for $^{19}$F in position 1; $\varphi$ is an azimuthal angle for ferromagnetic vector {\bf F} in basis plane. The formula (\ref{m9})  does provide a direct linkage between the   $^{19}$F  NMR frequencies and parameters of the crystalline $(p,c,x,l)$ and magnetic $(F,\varphi,\pm)$ structures. As of particular importance one should note a specific dependence of the   $^{19}$F  NMR frequencies on mutual orientation of the ferro- and antiferromagnetic vectors or the sign of the Dzyaloshinskii vector: upper signs in  (\ref{m9}) correspond to "right orientation"\, ($d(\theta)$\,>\,0) while lower signs do to "left orientation"\, ($d(\theta)$\,<\,0) as shown in Fig.\,\ref{fig10}.

\begin{figure}[t]
\centering
\includegraphics[width=7.5cm,angle=0]{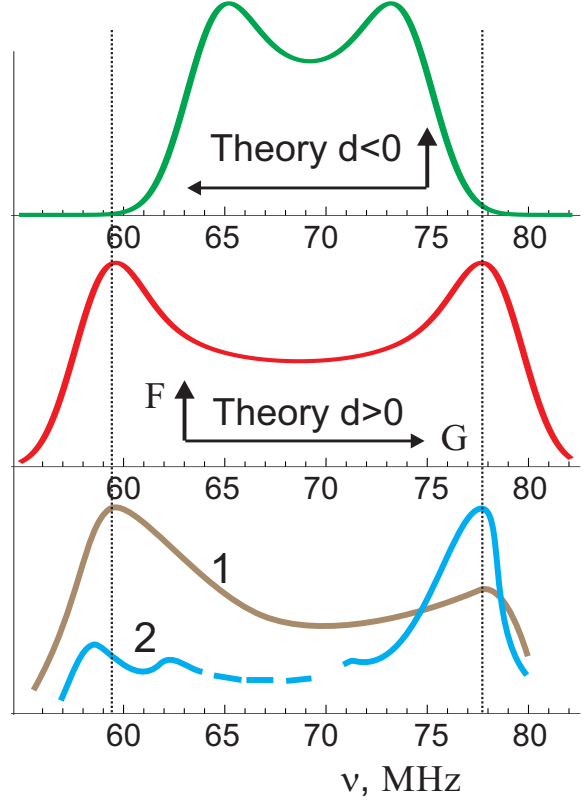}
\caption{(Color online) Comparison of simulated (upper panels) and experimental (bottom panel) zero-field $^{19}$F  NMR spectra in FeF$_3$.}
\label{fig_11}
\end{figure}	
	
For minimal and maximal values of the $^{19}$F NMR frequencies we have
\begin{eqnarray}{}
\label{m10}
\nu^{\pm}_{min}=\gamma_n[a_{xy}^2+a_{F}^2\pm 2a_Fa_{xy}F]^{1/2},\nonumber\\
\nu^{\pm}_{max}=\gamma_n[a_{xy}^2+a_{yz}^2+a_F^2F^2\mp 2a_Fa_{xy}F]^{1/2}.
\end{eqnarray}

Taking into account the smallness of isotropic HFI contribution, signs of $a_F$ and $A_{xy}$ we arrive at estimations
$$
\nu^{\pm}_{min}\simeq\gamma_n(|a_{xy}|\mp |a_FF|)=
2.92A_p\mp|a_FF|,
$$
$$
\nu^{\pm}_{max}\simeq\gamma_n\left([a_{xy}^2+a_{yz}^2]^{1/2}\pm
\frac{|a_{xy}}{[a_{xy}^2+a_{yz}^2]^{1/2}}|a_FF|\right)=
$$
\begin{equation}
3.65A_p\pm 0.8|a_FF|.
\end{equation}
Thus
\begin{equation}
\label{m12}
(\nu_{max}-\nu_{min})^{\pm}=0.68A_p\pm 1.8|a_FF|.
\end{equation}

By using the $A_s$ and $A_p$ values, typical for $^{19}$F
--Fe$^{3+}$ bonds\,\cite{Hall,Adam} we get (in MHz)
\begin{equation}
\label{m13}
\nu^+_{min}=57.6, \nu^+{max}=75.7, (\nu_{max}-\nu_{min})^+=18.1
\end{equation}
given "right hand side orientation" of {\bf F} and {\bf G} (Fig.\,\ref{fig10}) and
\begin{equation}
\label{m14}
\nu^-_{min}=61.4, \nu^-_{max}=72.7, (\nu_{max}-\nu_{min})^-=11.3
\end{equation}
given "left hand side orientation" of {\bf F} and {\bf G} (Fig.\,\ref{fig10}) .

 The zero-field $^{19}$F  NMR spectrum for single-crystalline samples of FeF$_3$ we simulated on assumption of negligibly small in-plane anisotropy\,\cite{Wolfe} is shown in Fig.\,\ref{fig_11} for two different mutual orientations of ${\bf F}$ and ${\bf G}$ vectors. For a comparison in Fig.\,\ref{fig_11} we adduce the experimental NMR spectra for polycrystalline samples of FeF$_3$\,\cite{Petrov,Zalesskii}, which are characterized by the same boundary frequencies despite rather varied shape. Obviously, the theoretically simulated NMR spectrum does nicely agree with the experimental ones only for "right" mutual orientations of ${\bf F}$ and ${\bf G}$ vectors, or $d(FeFe)$\,>\,0, in a full accordance with our theoretical sign predictions (see Table\,\ref{tablesign}).

The same result, $d(FeFe)$\,>\,0 follows from the the magnetic $x$-ray scattering amplitude measurements in the weak ferromagnet FeBO$_3$\,\cite{Dmitrienko}.

\subsection{The sign of the Dzyaloshinskii vector in FeBO$_3$ and $\alpha$-Fe$_2$O$_3$}
Making use of structural data for FeBO$_3$\,\cite{Diehl} we can calculate the $z$-component of the Dzyaloshinskii vector for Fe$_1$-O-Fe$_2$ pair, with Fe$_{1,2}$ in positions (1/2,1/2,1/2), (0,0,0), respectively, as follows:
$$
	d_z(12)=d_{12}(\theta )\left[{\bf r}_1\times{\bf r}_2\right]_z=+\frac{1}{3}(\frac{1}{2}-x_h)\frac{ab}{l^2}d_{12}(\theta )
$$
\begin{equation}
\approx +0.61\,d_{12}(\theta ) \, ,
\end{equation}
where $a$\,=\,4.626\,\AA, $b$\,=\,8.012\,\AA\, are parameters of the orthohexagonal unit cell, $x_h$\,=\,0.2981 oxygen parameter, $l$\,=\,2.028\,\AA\, is a mean Fe-O separation\,\cite{Diehl}.

Similarly to FeF$_3$ the DM energy per Fe$^{3+}$\,-\,O$^{2-}$\,-\,Fe$^{3+}$ bond can be written as follows
$$
	E_{DM}=d_z(12)\left[{\bf S}_1\times{\bf S}_2\right]_z = -2S^2d_z(12)(F_xG_y-F_yG_x)
$$
\begin{equation}
=+2\cdot 0.61\cdot S^2d_{12}(\theta )(F_xG_y-F_yG_x) \, .
\end{equation}
In other words, the "left" and "right" orientations of basic vectors are realized at $d(\theta )<0$ and $d(\theta )>0$, respectively.

Absolute magnitude of the ferromagnetic vector  equals numerically to an overt canting angle which can be found making use of familiar values of the Dzyaloshinskii field: $H_D\approx 100$\,kOe and exchange field: $H_E\approx 3.0\cdot
10^3$\,kOe\,\cite{Kotyuzhanskii,Diehl} as follows
\begin{equation}
F=H_D/2H_E\simeq 1.7\cdot 10^{-2}.
\end{equation}
If we know the Dzyaloshinskii field we can calculate the $d_{12}(\theta )$ parameter as follows
\begin{equation}
	H_D=\frac{6S}{g\mu_B}|d_z(12)|=\frac{6S}{g\mu_B}0.61|d(\theta )|=100\,kOe \, ,
\end{equation}
that yields $|d(\theta )|\cong$\,1.5\,K that is two times smaller than in YFeO$_3$. The difference can be easily explained, if one compares the superexchange bonding angles in FeBO$_3$ ($\theta \approx 125^{\circ}$) and YFeO$_3$ ($\theta \approx 145^{\circ}$), that is $cos\,\theta$(FeBO$_3$)/$cos\,\theta$(YFeO$_3$)$\approx$\,0.7, that makes the compensation effect of the $p$\,-\,$d$ and $s$\,-\,$d$ contributions to the $X$-factor (see Table\,\ref{tableXY}) more significant in borate than in orthoferrite. Interestingly that in their turn the structural factor $\left[{\bf r}_1\times{\bf r}_2\right]_z$ in FeBO$_3$ is 1.6 times larger than the mean value of the factor $\left[{\bf r}_1\times{\bf r}_2\right]_y$ in YFeO$_3$.

The sign of the Dzyaloshinskii vector in FeBO$_3$ has been experimentally found recently due to making use of a new technique based on interference of the magnetic x-ray scattering with forbidden quadrupole resonant scattering\,\cite{Dmitrienko}. The authors found that the the magnetic twist follows the twist in the intermediate oxygen atoms in the planes between the iron planes, that is the DM coupling induces a small left-hand twist of opposing spins of atoms at (0,0,0) and (1/2,1/2,1/2). This means that in our notations the Dzyaloshinskii vector for Fe$_1$-O-Fe$_2$ pair is directed along $c$-axis, $d_z(12)>$\,0, that is $d_{12}(\theta )>$\,0 in a full agreement with theoretical predictions (see Table\,\ref{tablesign}).



\section{Exchange-relativistic anisotropy: Unconventional features of conventional two-ion exchange anisotropy}
So called quasi-dipole two-ion exchange anisotropy (anisotropic exchange)
\begin{equation}
 V_{an} = \sum_{m,n,\alpha ,\beta}^{}K_{\alpha\beta}(mn)S_{m\alpha}S_{n\beta} \label{eq:spin}
 \end{equation}
with a traceless symmetric tensor $K_{\alpha\beta}(mn)$ of anisotropy parameters was introduced  by Van Vleck as early as in
1937\,\cite{VanVl}. For $S_{1}=S_{2}=1/2$ the anisotropy was considered in details by Moriya\,\cite{Moriya} and Yoshida\,\cite{Yosida}.
 Since then   the simple Hamiltonian  (\ref{eq:spin}) had been used increasingly without good reason for any 3$d$ ions and any spins $S\geq 1/2$. Simple square-law temperature dependence of the effective anisotropy constant $K_{TIA}(T) \sim B_{S}^{2}(T) \sim m^{2}(T)$
was addressed to be a "smoking gun" of the magneto-dipole or anisotropic exchange origin of the anisotropy (see, e.g., Refs.\,\cite{K-Cr2O3,Besser}).
However, a detailed many-electron analysis of the exchange-relativistic anisotropy to be a result of the third order perturbation contribution\,\cite{Nikiforov,TIA}
$$
 V_{an}(1,2) \sim \frac{V_{so}(1) V_{ex}(12) V_{so}(2)}{\Delta E^{2}} +
 \frac{V_{so}(1) V_{so}(2) V_{ex}(12)}{\Delta E^{2}} +
$$
 \begin{equation}
 \frac{V_{so}(1) V_{ex}(12) V_{so}(1)}{\Delta E^{2}}
+
\frac{V_{so}(1) V_{so}(1) V_{ex}(12)}{\Delta E^{2}}  \label{eq:van}
\end{equation}
(plus terms with $1\leftrightarrow 2$)
has revealed some novel features of the two-ion anisotropy missed in traditional approaches.
First of all, it concerns the tensor form of the anisotropic spin Hamiltonian. Simple quasi-dipole form (\ref{eq:spin}) is justified only for ions with $S_m=S_n$\,=\,1/2 and orbitally nondegenerate ground state, while for different spins the tensor form becomes more complicated. So, for the $S$-type ions, that is ions with orbitally nondegenerate ground state $A_{1g},A_{2g}$ in cubic crystal field (Cr$^{3+}$, Mn$^{2+}$, Fe$^{3+}$, Ni$^{2+}$,...) we arrive at an effective spin Hamiltonian as follows\,\cite{TIA}
\begin{equation}
 E_{an}=\sum_{k_{1} k_{2}}
    \rho_{k_{1}}\rho_{k_{2}}
  \left(
   K^{2}_{12}(k_{1},k_{2})\cdot
   \left[
    C^{k_{1}}( {\hat{\bf S}_{1}}) \times
    C^{k_{2}}( {\hat{\bf S}_{2}})
   \right]^{2}
  \right),
   \label{12}
\end{equation}
where does appear the tensor product of spherical tensorial harmonics,
$\rho_{k}(T)$ are temperature factors\,\cite{Callen}:
$$
 \rho_{0} = 1 ;\,\rho_{1} = B_{S}(T) =
   \frac{\left< S_{z}\right>}{S};
 \rho_{2} = \frac{\left< 3S_{z}^{2}-S(S+1)\right>}{S(2S-1)} ;
 $$
 \begin{equation}
 \rho_{3} = -\frac{\langle [3S(S+1)-1]\,{S_{z}}+S\,{S_{z}^{3}}\rangle}{S(S-1)(2S-1)},
\end{equation}
($\rho_{k}(T=0)$\,=\,1).

Along with a quasidipole term \((k_{1} = k_{2} =1)\) in \(V_{an}\) there is a number of novel nondipole terms with \(k_{1}k_{2} = 20(02),22\)  and \(k_{1}k_{2} = 13(31)\). Beyond that, $k_{1,2}$ should obey the triangle rule: $k_{1,2}\leq 2S_{m,n}$. It is worth noting that in addition to conventional spin-dependent exchange, the purely orbital spinless exchange interaction does contribute to the quasidipole exchange anisotropy\,\cite{TIA}.

\begin{figure}[t]
\centering
\includegraphics[width=8.5cm,angle=0]{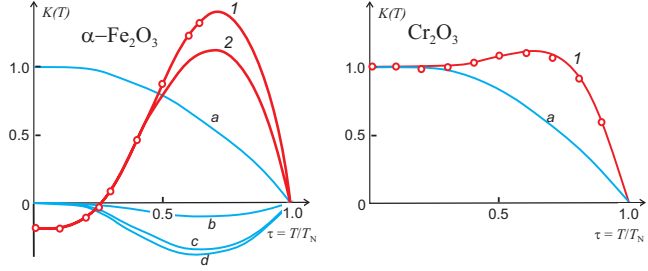}
\caption{(Color online) The temperature dependence of effective anisotropy constants in $\alpha$-Fe$_2$O$_3$ (left) and Cr$_2$O$_3$ (right). Circles are experimental data from Refs.\,\cite{K-Fe2O3} and \,\cite{K-Cr2O3} for $\alpha$-Fe$_2$O$_3$ and Cr$_2$O$_3$, respectively.
Curves 1 represent the result of the fitting with the four-parameter formula (\ref{K(T)}), curve 2 for hematite shows  the result of the fitting that implies only conventional  contributions of single-ion and quasi-dipole anisotropy.
Curves $a$, $b$, $c$, and $d$ represent the temperature dependence of the quasi-dipole contribution ($a$) and of the "non-dipole"\, factors $(\rho_{2}-\rho_{1}^{2})$, $(\rho_{2}^2-\rho_{1}^{2})$, and $(\rho_{1}\rho_{3}-\rho_{1}^{2}))$, respectively.} \label{fig11}
\label{fig12}
\end{figure}

Within mean-field approximation  the temperature dependence of effective 2nd order exchange-relativistic anisotropy constant(s) for magnets with equivalent on-site spins can be represented as follows\,\cite{TIA}:
$$
 K(T) = K(0) \rho_{1}^{2} + K_{20}(\rho_{2} - \rho_{1}^{2}) +
        K_{22}(\rho_{2}^{2} - \rho_{1}^{2}) +
 $$
  \begin{equation}
        K_{13}(\rho_{1}\rho_{3}-\rho_{1}^{2}), \label{K(T)}
\end{equation}
where the temperature factors $(\rho_{2}-\rho_{1}^{2})$, $(\rho_{2}^2-\rho_{1}^{2})$, and $(\rho_{1}\rho_{3}-\rho_{1}^{2}))$
turn into zero both at T\,=\,0\,K and T\,=\,T$_{N}$(T$_{c}$).
The constant $K_{11}$ for conventional quasi-dipole anisotropy is determined as $K_{11} = K(0) - K_{20} - K_{22} - K_{13}$.
It is worth noting that the addition of the magneto-dipole and single-ion anisotropies gives rise only to a renormalization of the $K(0)$ and  $K_{20}$ constants, respectively, so that the expression (\ref{K(T)}) is believed to be an universal four-parametric formula for the temperature dependence of the 2nd order anisotropy constants. As we see in Fig.\,\ref{fig12} the formula allows us to nicely describe nontrivial temperature dependence of effective anisotropy constants in $\alpha$-Fe$_2$O$_3$ and Cr$_2$O$_3$. Later, this approach was used to describe the temperature dependence of the anisotropy constants in YFeO$_3$\,\cite{Egoyan}.

\section{Antisymmetric supertransferred hyperfine interaction as electron-nuclear counterpart of DM coupling }

 Detailed analysis of the $^{57}$Fe NMR data in orthoferrites\,\cite{Luetgemeier} allowed us to reveal  antisymmetric supertransferred hyperfine (ASTHF) coupling
\begin{equation}
	{\hat H}_{ASTHF}=\sum_{m>n}({\bf a}_{mn}\cdot[{\bf I}_m\times{\bf S}_n])
	\label{ASTHF}
\end{equation}
as an electron-nuclear analogue of the DM antisymmetric exchange and to determine its contribution to the local field: $H_{ASTHF}\approx$\,0.26\,T as compared with corresponding isotropic contribution of 5.8\,T\,\cite{Moskvin-ASTHF}.
Here ${\bf I}$ is a nuclear spin, ${\bf a}_{mn}$ electron-nuclear analogue of the Dzyaloshinskii vector.

For the first time such a electron-nuclear coupling was considered by Ozhogin\,\cite{Ozhogin-ASTHF}, the microscopic theory was considered by Moskvin\,\cite{Moskvin-ASTHF}. Furthermore, in Ref.\,\cite{Moskvin-ASTHF} we have shown that the experimentally known external field dependencies of the $^{57}$Fe NMR in the orthoferrites\,\cite{Luetgemeier} do allow us to find out and estimate the ASTHF coupling.

Indeed, taking into account the four-sublattice magnetic structure of the orthoferrite RFeO$_{3}$ with nonmagnetic R-ions (La, Y, Lu) the local field on the $^{57}$Fe$_{i}$ nuclei in one of the $4b$-sites can be written as follows
\begin{equation}
{\bf H}_{loc}(i)=a_{G}(i){\bf G}+a_{F}(i){\bf F}+a_{C}(i){\bf C}+a_{A}(i){\bf A}+{\bf \stackrel{\leftrightarrow}{a}}(i){\bf G}
\label{art3}
\end{equation}
where ${\bf G},~{\bf F},~{\bf C},~{\bf A}$ are the  basis magnetic vectors normalized as follows: ${\bf G}^{2}+{\bf F}^{2}+{\bf C}^{2}+{\bf A}^{2}=1$. Here, the first four terms represent the contribution of the dominant isotropic on-site  and supertransferred inter-site hyperfine interactions,  while  the last term does the contribution of the anisotropic hyperfine interactions. Hereafter we take into account  that  F, C, A $\approx 10^{-2}$G, and assume that the anisotropic contribution does not exceed  values on the order of $\propto 1\%$ of main isotropic contribution $H_{0}=a_{G}G$.

The external field dependence of the $^{57}$Fe NMR frequency for the $\Gamma_{4}(G_{x},A_{y},F_{z})$ magnetic configuration is as follows:
\begin{equation}
\nu_{\Gamma_{4}}({\bf h}\parallel{\bf c})=1-a_{xx}+(a_{zx}G_{x}+a_{F}F_{z})h+h^{2}/2,
\label{art4}
\end{equation}
where we make use all the quantities $a_{zx},~a_{F},~a_{xx},~h$ in units of $H_{0}$ (in YFeO$_{3}$ given T\,=\,4.2\,K $H_{0}$\,=\,551\,kOe\,\cite{Luetgemeier}), while $\nu$ is in units of $\nu_{0}=\gamma H_{0}/2\pi$ ($\gamma/2\pi=0.138$\,MHz/kOe). The field derivative $(\partial\nu/\partial h)_{h=0}$
\begin{equation}
[\partial\nu_{\Gamma_{4}}({\bf h}\parallel{\bf c})/\partial h]_{h=0}=a_{F}F_{z}+a_{zx}G_{x}\, ,
\label{art5}
\end{equation}
is a sum of a ferromagnetic($a_{F}F_{z}$) and antiferromagnetic ($a_{zx}G_{x}$) contributions, respectively. Application of the magnetic field parallel to $a$-axis (${\bf h}\parallel {\bf a}$) does induce a spin-reorientational transition $\Gamma_{4}(G_{x},A_{y},F_{z})-\Gamma_{2}(F_{x},C_{y},G_{z})$ so that for angular $\Gamma_{42}$ configuration
$$
\nu_{\Gamma_{42}}({\bf h}\parallel{\bf a})=1-(a_{xx}G_{x}^{2}\pm 2a_{zx}^{(s)}G_{x}G_{z}+a_{zz}G_{z}^{2})+
$$
\begin{equation}
(a_{xz}G_{z}+a_{F}F_{x}\pm a_{G}G_{x})h+h^{2}/2
\label{art6}
\end{equation}
where $a_{zx}^{(s)}$ is a symmetrical part of
 $a_{zx}$: $a_{zx}^{(s)}=(a_{zx}+a_{xz})/2$). The signs $\pm$ in (\ref{art6}) correspond to nuclei in positions 1,3 and 2,4, respectively (see Ref.\,\cite{Karnachev}). The $\Gamma_4$\,-\,$\Gamma_2$ spin reorientation is accompanied by the $^{57}$Fe NMR frequencies splitting whose magnitude
\begin{equation}
\Delta\nu=2(2a_{zx}^{s}G_{z}+h)G_{x}
\label{art7}
\end{equation}
allows us to find out the $a_{zx}^{(s)}$ parameter, or, strictly speaking, its absolute value: $|a_{zx}^{(s)}|=3.2\cdot 10^{-3}$ in YFeO$_{3}$\,\cite{Luetgemeier},  $|a_{zx}^{(s)}|=3.4\cdot 10^{-3}$ in ErFeO$_{3}$, and $|a_{zx}^{(s)}|=2.9\cdot 10^{-3}$ in HoFeO$_{3}$\,\cite{Karnachev}.

Experimental value of the field derivative $[\partial\nu_{\Gamma_{4}}({\bf h}\parallel{\bf c})/\partial h]_{h=0}$ \,=\,-10.2$\cdot 10^{-3}$ in YFeO$_3$\,\cite{Luetgemeier} with taking account of  $a_{F}=2H_{STHF}/H_{0}-1$\,=\,-0.79 ($H_{STHF}$ is the contribution of the STHF interaction $^{57}$Fe--O$^{2-}$--Fe$^{3+}$ to the local field)\,\cite{Luetgemeier} and $F_{z}=1.1\cdot 10^{-2}$\,\cite{Jacobs} allows us to find out the value $a_{zx}G_x=-1.6\cdot10^{-3}$.
Finally we obtain that for $a_{zx}^{(s)}=\pm 3.2\cdot 10^{-3}$, $a_{zx}^{(a)}=\mp 1.6\cdot 10^{-3}$ given $F_z>0, G_x<0$ ($G_x\approx$\,-1) and $a_{zx}^{(a)}=\mp 4.8\cdot 10^{-3}$ given $F_z>0, G_x>0$ ($G_x\approx$\,+1), that is for $d(\theta)>0$ and $d(\theta)<0$, respectively. In any case antisymmetric and symmetric parts of the anisotropic hyperfine interaction in YFeO$_3$ are of a comparable magnitude. The origin of the antisymmetric part  ${\bf \stackrel{\leftrightarrow}{a}}^{a}$ can be related only with antisymmetric STHF interaction ${\hat H}_{ASTHF}$ (\ref{ASTHF}) that is with electron-nuclear analogue of the DM coupling. Then
\begin{equation}
a_{zx}^{(a)}(i)=-\frac{S}{g_{n}\beta_{n}}\sum_{j}a_{y}(ij).
\label{art8}
\end{equation}
The ASTHF interaction is a result of a combined action of the generalized STHF interaction $^{57}$Fe--O$^{2-}$--Fe$^{3+}$\,\cite{Kolya} and spin-orbital coupling for Fe$^{3+}$ ion. As for conventional spin-spin DM coupling the electron-nuclear analogue of the Dzyaloshinskii vector depends on the superexchange   $^{57}$Fe--O$^{2-}$--Fe$^{3+}$ bond geometry
\begin{equation}
{\bf a}(ij)=a(\theta)\left[{\bf r}_{i}\times{\bf r}_{j}\right],
\label{art9}
\end{equation}
where ${\bf r}_{i},~{\bf r}_{j}$ are unit cation-anion radius vectors, and
\begin{equation}
a(\theta)=a_{1}+a_{2}\cos\theta,
\label{art10}
\end{equation}
where $\theta$ is the cation-anion-cation bond angle.

For a rough estimate of the $a$ parameter one may use relation $a/A_{STHF}\leq\xi/\Delta E$ where $\xi$ is a single electron spin-orbital coupling parameter for 3$d$ electron; $\Delta E$ is the energy of the excited terms of the $^{4}T_{1}$ type for Fe$^{3+}$ ion; $A_{STHF}$ is a isotropic STHF constant:
\begin{equation}
\hat{V}_{STHF}=\sum_{i\neq j}A_{STHF}(ij)({\bf I}_{i}\cdot{\bf S}_{j}).
\label{art11}
\end{equation}
In our case $\xi\leq 5\cdot10^{2}~cm^{-1}$,  $\Delta E\geq 10^{4}~cm^{-1}$, and we arrive at
$$a/A_{STHF}\leq 5\cdot 10^{-2}$$ that nicely agrees with estimate based on the experimental data\,\cite{Luetgemeier}
$$
|a_{zx}^{(a)}/H_{STHF}|\approx 4.6\cdot 10^{-2} \, .
$$
 The ratio is comparable with the ratio of the Dzyaloshinskii field
$H_{D}$ to the exchange field $H_{E}$: in YFeO$_{3}$ $H_{D}/H_{E}\approx 2.2\cdot 10^{-2}$. All that is quite natural as the Dzyaloshinskii field is of the exchange-relativistic nature: $H_{D}/H_{E}\propto\xi/\Delta E$, hence, $|a_{zx}^{(a)}/H_{STHF}|\propto H_{D}/H_{E}$. In other words, if $H_{STHF}$ is an electron-nuclear analogue of the exchange field, the $H_{ASTHF}=|a_{zx}^{(a)}|$ is an electron-nuclear analogue of the Dzyaloshinskii field. In YFeO$_{3}$
$H_{STHF}$\,=\,58\,kOe, $H_{ASTHF}$\,=\,2.6\,kOe given $G_x>0$ or $H_{ASTHF}$\,=\,0.9\,kOe given $G_x<0$.
For rough estimate of the electron-nuclear Dzyaloshinskii field one may use $H_{ASTHF}\approx (H_{D}/H_{E})H_{STHF}$.

Antisymmetric STHF interaction should be observed in other weak ferromagnets. It is worth noting that for an easy-plane phase of rhombohedral weak ferromagnets such as FeBO$_{3}$, FeF$_{3}$, $\alpha$-Fe$_{2}$O$_{3}$, the antiferromagnetic contribution to the field derivative $[\partial\nu({\bf h}\bot{\bf C}_{3})/\partial h]_{h=0}$ is determined only by the ASTHF interaction:
\begin{equation}
[\partial\nu({\bf h}\bot{\bf C}_{3})/\partial h]_{h=0}=a_{xy}^{(a)}G_{y}+a_{F}F_{x},
\label{art12}
\end{equation}
that makes its detection and estimation more easier than in orthoferrites.

It should be noted that the electron-nuclear double resonance (ENDOR) measurements in Pb$_5$Ge$_3$O$_{11}$:Gd$^{3+}$ revealed antisymmetric $^{207}$Pb--O$^{2-}$--Gd$^{3+}$ supertransferred hyperfine interaction\,\cite{Rokeah} whose origin can be related with the ligand spin-orbital contribution.

\section{Antisymmetric exchange-relativistic spin-other-orbit coupling and unconventional magnetooptics of weak ferromagnets}

Interestingly that circular magnetooptic effects in weak ferromagnets are anomalously large and are comparable with the effects in ferrite garnets despite two-three orders of magnitude smaller magnetization\,\cite{Kahn,Tabor,Chetkin}. In 1989 the anomaly has been assigned to a novel type of magnetooptical mechanisms related with so called spin-other-orbit coupling\,\cite{MO-Pisarev}.

Combined effect of a conventional on-site spin-orbital coupling and orbitally off-diagonal exchange coupling for an excited orbitally degenerated state can give rise to a novel type of exchange-relativistic interaction, so called spin-other-orbit coupling, whose bilinear form can be written as a sum of isotropic, anisotropic antisymmetric, and anisotropic symmetric terms, respectively
$$
{\hat V}_{SoO}=\sum_{m>n}\lambda^{(0)}_{mn}({\bf L}_{m}\cdot{\bf S}_{n})+\sum_{m>n}(\pmb{\lambda}_{mn}\cdot[{\bf L}_{m}\times{\bf S}_{n}])+
$$
\begin{equation}
\sum_{m>n} ({\bf L}_{m}\stackrel{\leftrightarrow}{\pmb{\lambda}}_{mn}{\bf S}_{n}) \, .
\label{SoO}
\end{equation}
It is worth noting that $\pmb{\lambda}_{mn}$ has the symmetry of the Dzyaloshinskii vector, while the last term  has the symmetry of the two-ion quasidipole spin anisotropy.   Generally speaking, all the three terms can be of a comparable magnitude.

Interestingly, the contribution to the bilinear interaction ${\hat V}_{SoO}$ is made by both the spin-dependent
exchange and spin-independent purely
orbital exchange. However, the spin-dependent exchange leads to
the occurrence of additional nonlinear spin-quadratic
terms, the contribution of which can be taken into
account by the formal replacement of the linear spin
operator ${\bf S}_n$ in (18) for the nonlinear operator  ${\bf S}_{mn}$
\begin{eqnarray}
\hat S_q(mn)\,=\, \hat S_q(n) \,+\,
\gamma \,\left[\hat V^2 \Bigl( S(m) \Bigr) \, \times \, S^1(n)\right]^1_q \, = \nonumber \\
= \hat S_q(n)+ \gamma  \sum_{q_1,q_2} \left[
\begin{array}{ccc}
2 & 1 & 1 \\
q_1 & q_2 & q
\end{array}\right] \hat V^2_{q_1} \Bigl( S(m) \Bigr) S_{q_2}(n)\>,
\label{Smn}
\end{eqnarray}
where $\; V^2_q(S)\;$  is the rank 2 spin irreducible tensor operator. In particular,
\begin{equation}
\hat V_0^2(S)\,=\, 2\,\left[\frac{(2S\,-\,2)!}{(2S\,+\,3)!}\right]^{1/2}
\Bigl( 3 \hat S_z^2 \>-\> S(S \,+\,1) \Bigr)\>.   \label{V2}
\end{equation}
The coefficient $\; \gamma \;$ in (\ref{Smn}) can be calculated for specific terms.
The isotropic part of $\;V_{so}^{ex}\;$ can be presented, in the
general case, as follows
$$
V_{so}^{ex} \>= \> \sum_{mn}\, \lambda (mn) \left({\bf L}(m) \cdot {\bf S}(n)\right) \,+
$$
\begin{equation}
+\sum_{m \not= n}\, \lambda ^{'}(mn) \Bigl( {\bf L}(m) \cdot {\bf S}(m) \Bigr)
\Bigl( {\bf S}(m) \cdot {\bf S}(n) \Bigr)\>.\label{eq:30}
\end{equation}
Similarly to the Dzyaloshinskii vector, to estimate the
parameters of the spin–other orbit coupling, we can
use the simple relation
\begin{equation}
\> \lambda (mn) \> \approx \> \frac{\lambda ^{'} I^{'}}
{\Delta E_{S \Gamma}}\> ,\label{eq:31}
\end{equation}
where $\lambda^{'}\,$ and $\,I^{'}\,$ are the spin–orbital constant for the $T_1$-, $T_2$-states and the nondiagonal exchange parameter, respectively,
 $\Delta E_{S \Gamma}$ is a certain excitation energy. The simple
estimation shows that, due to ${\hat V}_{SoO}$, the effective magnetic
fields acting on the $T_1$ and $T_2$ orbital states, e.g.,
for Fe$^{3+}$ions in ferrites, can amount about 100\,T and more.


We have shown that an antisymmetric exchange-relativistic spin-other-orbit coupling
gives rise to an unconventional "antiferromagnetic" contribution to the circular magnetooptics for weak ferromagnets which can surpass conventional "ferromagnetic" term\,\cite{MO-Pisarev,MO} (see, also Ref.\,\cite{Zubov}).

The circular magnetooptics is governed by an axial gyration vector ${\bf g}$, which is  dual to the permittivity tensor $\varepsilon_{ij}$. For instance the Faraday rotation $\Theta_{F}$ in noncubic crystals can be written as follows
\begin{equation}
\Theta_{F}=A({\bf g}\cdot{\bf n})	\, ,
\end{equation}
where ${\bf n}$ is a unit vector in the direction of light propagation
${\bf k}$, and $A$ is a coefficient which depends on the direction of ${\bf k}$, on the polarization of light, and on the principal values of the refractive index. The gyration vector has the same symmetry properties as the ferromagnetic moment that does justify a well known relation
\begin{equation}
	{\bf g}=\stackrel{\leftrightarrow}{\pmb{\alpha}}{\bf F}+ \stackrel{\leftrightarrow}{\pmb{\gamma}}{\bf H}_{ext} \, ,
\end{equation}
where the  gyration vector is a sum of so-called ferromagnetic and diamagnetic terms, respectively.
However, in weak ferromagnets where orthogonal components of the ferromagnetic and antiferromagnetic vectors may transform identically we arrive at an additional "antiferromagnetic" contribution. For example, in the case of orthoferrites this term, if to neglect weak antiferromagnetic modes, can be written as follows:
\begin{equation}
	\Delta{\bf g}=\stackrel{\leftrightarrow}{\pmb{\beta}}{\bf G}
\end{equation}
 with the only nonzero, and generally speaking unequal, components $\beta_{zx}$ and $\beta_{xz}$ of the $\stackrel{\leftrightarrow}{\pmb{\beta}}$ tensor. Despite for main isotropic contribution the components of the $\stackrel{\leftrightarrow}{\pmb{\alpha}}$ tensor are expectedly significantly larger than the   $\stackrel{\leftrightarrow}{\pmb{\beta}}$ tensor components, the relation $F\ll G$ which is typical for weak ferromagnets points to a puzzling effect of a probably leading antiferromagnetic contribution to the gyration vector.

It is generally accepted that the major (if not dominant) role in the magnetooptic rotation of visible and ultraviolet
light in rare-earth orthoferrites is played by allowed electric-dipole O2$p$--Fe3$d$ charge transfer ${}^6A_{1g}-{}^6T_{1u}$ transitions in octahedral
complexes FeO$_6$\,\cite{Kahn,MO,Pisarev}, in particular, due to an orbital Zeeman splitting of the excited  ${}^6T_{1u}$ term with effective orbital moment $L$\,=\,1. In addition to conventional orbital Zeeman  $V_Z^{orb}=-\beta_e \sum_{m}({\bf L}_m\cdot {\bf H}_{ext})$ and on-site spin-orbital $V_{SO}=\lambda\sum_{m}({\bf L}_{m}\cdot {\bf S}_m)$ couplings such a splitting is caused by unconventional spin-other-orbit coupling (\ref{SoO}). In all the cases we deal with a real or effective orbital magnetic field.
The contribution of the isolated ${}^6T_{1u}$ term to the gyration vector can be represented as follows\,\cite{MO-Pisarev,MO}:
$$
	{\bf g}=\left(\frac{n_0^2+2}{3}\right)^2\frac{2\pi e^2f_{AT}}{m\omega_0}\frac{\partial F(\omega ,\omega_0)}{\partial\omega_0}
$$	
$$
	(-N\beta_e{\bf H}_{ext}+\lambda_{eff}\sum_m\langle{\bf S}_{m}\rangle +\sum_{m>n}\lambda^{(0)}_{mn}\langle{\bf S}_{n}\rangle -
$$
\begin{equation}
  \sum_{m>n}[\pmb{\lambda}_{mn}\times\langle{\bf S}_{n}\rangle ] +\sum_{m>n} \stackrel{\leftrightarrow}{\pmb{\lambda}}_{mn}\langle{\bf S}_{n}\rangle    )\, ,
	\label{gg}
\end{equation}
where $N$ is the number of FeO$_6$ clusters in the unit volume, $f_{AT}$ and $\hbar\omega_0$ are the oscillator strength and the energy of the ${}^6A_{1g}-{}^6T_{1u}$ transition, respectively, $F(\omega ,\omega_0)$ is the dispersion factor. Here, the first and second terms determine conventional "local" diamagnetic and isotropic ferromagnetic contributions, respectively, while the three other  terms derived from the spin-other-orbit interaction determine unconventional  "nonlocal"  contribution, though the first one gives a simple correction to the ferromagnetic term. However, the second and third nonlocal terms provide novel antisymmetric and symmetric anisotropic antiferromagnetic contributions to gyration vector, respectively. Their effect was experimentally studied in orthoferrite YFeO$_3$\,\cite{MO-Pisarev}. An analysis of the dependence $\Theta_F({\bf H}_{ext})$  made it possible to determine all the contributions to the gyration vector ($\lambda$\,=\,0.6328\,$\mu$m):
$$
	\alpha_{zz}F_z=(0.95\pm 0.55)\cdot 10^{-3};\, \beta_{zx}G_x=(3.15\pm 0.55) )\cdot 10^{-3};\,
	$$
	$$
\alpha_{xx}F_x=(0.2\pm 0.7)\cdot 10^{-3};\, \beta_{xz}G_z=(-2.1\pm 1.0) )\cdot 10^{-3};\,
$$
\begin{equation}
\gamma_{zz}\approx \gamma_{xx}=(-1.1\pm 2.8) \cdot 10^{-6}\,kOe^{-1}\,  .
\end{equation}
Interestingly, rather large measurement errors allow for certain to determine only the fact of a large if not a dominant antisymmetric antiferromagnetic contribution related with antisymmetric spin-other-orbit coupling.

Existence of spontaneous spin-reorientational phase transitions $\Gamma_4(F_zG_x)\rightarrow\Gamma_2(F_xG_z)$ in several rare-earth orthoferrites does provide large opportunities to study anisotropy of circular magnetooptics\,\cite{Kahn,Tabor,Chetkin,MO}. Gan'shina {\it et al.} measured the equatorial Kerr effect in TmFeO$_3$ and HoFeO$_3$ and have found the the gyration vector anisotropy in a wide spectral range 1.5-4.5\,eV\,\cite{MO}. The magnetooptical spectra were nicely fitted within a microscopic model theory based on the dominating contribution of the O2$p$--Fe3$d$ charge transfer transitions and spin-other-orbit coupling  in FeO$_6^{9-}$ octahedra. Studies have demonstrated a leading contribution of the antisymmetric spin-other-orbit coupling and allowed to estimate effective orbital magnetic fields in excited $^6T_{1u}$ states of the FeO$_6^{9-}$ octahedra, $H_L\sim 100\,T$. These anomalously large fields can be naturally explained to be a result of strong exchange interactions of the charge transfer $^6T_{1u}$ states with nearby octahedra that are determined by a direct $p$\,-\,$d$ exchange.
Whereas the existence of the antiferromagnetic contribution to the gyration vector is typical of a large number of multisublattice magnetic
materials, the antisymmetry of the tensor $\stackrel{\leftrightarrow}{\pmb{\beta}}$  is a specific feature
of weak ferromagnets alone. In the case of rhombohedral weak ferromagnets such as FeBO$_3$, FeF$_3$, or $\alpha$-Fe$_2$O$_3$,
the tensor $\stackrel{\leftrightarrow}{\pmb{\beta}}$, governing the antiferromagnetic contribution
to the Faraday effect is entirely due to the antisymmetric
contribution, in view of the requirements imposed by the
crystal symmetry. In crystals of this kind the appearance of
the antiferromagnetic contribution to the gyration vector is
entirely due to allowance for the antisymmetric spin-other-orbit coupling.

\section{Antisymmetric exchange-relativistic spin-dependent electric polarization}

I.~E. Dzyaloshinskii in 1959 theoretically predicted the existence of the magnetoelectric effect (ME) in antiferromagnetic Cr$_2$O$_3$\,\cite{Dzyalo_ME}, and a year later, D.~N. Astrov recorded the magnetization induced by an electric field\,\cite{Astrov}. Since the prediction and discovery of the ME effect in Cr$_2$O$_3$, several different mechanisms of magnetoelectric coupling have been proposed\,\cite{Fiebig}, but a real breakthrough in this direction is associated with the discovery and studying of multiferroics.

Currently  two essentially different spin structures of net electric polarization in crystals are considered: i) a bilinear {\it nonrelativistic symmetric} spin coupling\,\cite{TMS}
\begin{equation}
	{\bf P}_s=\sum_{mn}{\bf \Pi}_{mn}^s({\bf S}_m\cdot {\bf S}_n)\,
	\label{TMS}
\end{equation}
or ii) a bilinear {\it relativistic antisymmetric} spin coupling\cite{Katsura,Sergienko}
\begin{equation}
\hat {\bf P}_a=\sum_{m>n} {\bf \stackrel{\leftrightarrow}{\Pi}}_{mn} [{\bf S}_m\times{\bf S}_n])\, \,	
\end{equation}
respectively. Strictly speaking, ${\bf \stackrel{\leftrightarrow}{\Pi}}_{mn}$ is a rank-2 tensor, whose antisymmetric component yields the contribution to the electric polarization ${\bf P}_a$ as follows
\begin{equation}
	{\bf P}_a=\sum_{mn}\left[{\bf \Pi}_{mn}^a\times \left[{\bf S}_m\times {\bf S}_n\right]\right]\, ,
\end{equation}
The effective dipole moments ${\bf \Pi}_{mn}^{s,a}$  depend on the $m,n$ orbital states and the $mn$ bonding geometry.

If the first term  stems somehow or other from a spin isotropic Heisenberg exchange interaction (see, e.g. Refs.\,\cite{TMS,Druzhinin}), the second term does from antisymmetric Dzyaloshinskii-Moriya (DM) coupling. Following Katsura {\it et al.}\,\cite{KNB} the electric dipole ${\bf P}_a$ is said to be induced by a spin current mechanism, since the vector product $\left[{\bf S}_m\times {\bf S}_n\right]$ is proportional to the bond spin current, where the Dzyaloshinskii vector ${\bf d}_{mn}$ acts as its vector potential.
Namely the second, or "spin-current"\, term is frequently considered to be one of main contributors to multiferroicity\,\cite{Cheong}, however, at present there is no reliable theoretical justifications and experimental evidences for its dominating over conventional symmetric isotropic term\,\cite{SLD,multiferro}.

Microscopic quantum theory of ME effect has not yet been fully developed, although several scenarios for particular materials have been proposed. Katsura {\it et al.}\,\cite{KNB}  presented  a  mechanism of the giant ME effect theoretically derived "in terms of a microscopic electronic model for noncollinear magnets". The authors derived the expression for the electric dipole moment for the spin pair as follows:
\begin{equation}
	{\bf P}_{ij}=a \left[ {\bf R}_{ij}\times \left[{\bf S}_{i}\times {\bf S}_{j}\right]\right]\, ,
	\label{Katsura}
\end{equation}
where ${\bf R}_{ij}$ denotes the vector connecting the two sites $i$ and $j$, ${\bf S}_{i,j}$ are spin moments, $a$ is an exchange-relativistic parameter.
However, the original "spin-current" model by Katsura {\it et al.}\,\cite{KNB}  seems to be  questionable  as the authors proceed with an unrealistic scenario\,\cite{multiferro}.

The spin-current model can explain direction of ferroelectric polarization for P$_{bmn}$ spin-cycloidal perovskite manganites, however, cannot explain the polarization anisotropy in spin-spiral LiCu$_2$O$_2$ and LiVCuO$_4$, the direction of ferroelectric polarization for spin-cycloilal delafossites, such as AgFeO$_2$ and $\alpha$-NaFeO$_2$, the absense of polarization in spin-spiral NaCu$_2$O$_2$.  It cannot explain an emergence of ferroelectricity associated with proper crew magnetic ordering in several multiferroics, including CuFeO$_2$, CuCrO$_2$, AgCrO$_2$, Cu$_3$Nb$_2$O$_8$, CaMn$_7$O$_{12}$, and RbFe(MoO$_4$)$_2$, because the propagation vector ${\bf R}_{ij}\parallel [{\bf S}_i\times{\bf S}_j]$. In other words, for a most part of multiferroics the spin-current model does not work.

Alternative mechanism of giant magnetoelectricity in the perovskite manganites  based on the antisymmetric DM type magnetoelastic coupling was proposed by Sergienko and  Dagotto\,\cite{Dagotto}. The authors took into account strong dependence of the Dzyaloshinskii vector  on the superexchange bond angle and the displacement of the intermediate ligand.
However, here we meet  with a "weak" contributor. Indeed, the minimal value of $\gamma$ parameter ($\gamma = d{\bf D}/d{\bf R}$) needed to explain experimental  phase transition in multiferroic manganites is two orders of magnitude larger than the reasonable microscopic estimations\,\cite{Dagotto}.

Size of the macroscopic polarization {\bf P} in nonmagnetic ferroelectrics computed by modern {\it ab-initio} band structure methods agrees exceptionally well with the ones observed experimentally. However, state of the art {\it ab-initio} computations for different multiferroics: manganites HoMnO$_3$,  TbMn$_2$O$_5$, HoMn$_2$O$_5$, spin spiral chain cuprates \LiV \, and \Li \, yield data spread within one-two orders of magnitude with  absolutely ambiguous and unreasonable values of polarisation.  Indeed, the basic starting points of the current versions of such spin-polarized approaches as the LSDA exclude any possibility to obtain a reliable quantitative estimation of  the spin-dependent electric polarization in multiferroics. Basic drawback of the spin-polarized approaches is that these start with a  local density functional which implies  presence of
a large fictious local {\it one-electron} spin-magnetic field. Magnitude of the field is considered to be  governed by the intra-atomic Hund exchange, while its orientation does by the effective molecular, or inter-atomic exchange fields. Despite the supposedly spin nature of the field it produces an unphysically giant spin-dependent rearrangement of  the charge density that cannot be reproduced within any conventional technique operating with spin Hamiltonians.
In such a case the straightforward application of the LSDA scheme can lead to an unphysical overestimation of the effects or even to qualitatively incorrect results due to an unphysical effect of a breaking of spatial symmetry induced by a spin configuration.

Overall,   the LSDA approach seems to be more or less  justified for a semiquantitative description of exchange coupling effects for  materials with a classical N\'eel-like collinear magnetic order. However, it can lead to erroneous results for systems and effects where the symmetry breaking and quantum fluctuations are of a principal importance such as:  i) noncollinear spin configurations, in particular,  in quantum s=1/2 magnets, ii)  relativistic effects, such as the symmetric spin anisotropy, antisymmetric DM coupling, and, iii) spin-dependent electric polarization. Indeed, a correct treatment of these high-order perturbation effects needs in a correct account both of local symmetry and of quantum fluctuations (see, e.g., Ref.\,\cite{JETP-2007}).

A systematic standard microscopic theory of spin-dependent electric polarization which implies the derivation of effective spin operators for  nonrelativistic and relativistic contributions to electric polarization of the generic three-site two-hole cluster such as Cu$_1$\,-\,O\,-\,Cu$_2$ has been proposed in Ref.\,\cite{SLD,multiferro}. The authors made use of conventional well-known approaches to account for the $p$\,-\,$d$ covalent effects, intra-atomic correlations, crystal field, and spin-orbital coupling. Despite the description was focused on a three-site Cu$_1$\,-\,O\,-\,Cu$_2$ two-hole system typical for cuprates with a tetragonal Cu on-site symmetry and Cu3$d_{x2-y2}$   ground states, the generalization of the results on the M$_1$\,-\,O\,-\,M$_2$ clusters in other 3$d$ oxides seems to be a trivial procedure.

The effective electric polarization differs for the singlet and triplet pairing due to a respective singlet-triplet difference in the hybridization amplitudes. Hence we may introduce an effective nonrelativistic {\it exchange-dipole} spin operator
\beq
{\bf \hat P}_{12}= {\bf \hat P}_{12}^{(0)}+ {\bf \Pi}_{12}({\bf \hat s}_1\cdot {\bf \hat s}_2)
\label{non}
\eeq
 with an {\it exchange-dipole} moment
\beq
{\bf \Pi}_{12}=\langle {\bf P}\rangle _{S=1}-\langle {\bf P}\rangle _{S=0}\,
\eeq
and a spinless contribution ${\bf \hat P}_{12}^{(0)}=\frac{1}{4}\left(3\langle {\bf P}\rangle _{S=1}+\langle {\bf P}\rangle _{S=0} \right)$.
 It is worth noting that the net local  electric polarization lies in the Cu$_1$\,-\,O\,-\,Cu$_2$ plane.
As it was shown in Ref.\,\cite{SLD} (see, also Refs.\,\cite{multiferro}), in general, the {\it exchange-dipole} moment can be written as a superposition of the “longitudinal” and “transversal” contributions as follows:
\begin{equation}
	{\bf \Pi}_{12}=p_{\parallel}{\bf R}_{12}+p_{\perp}{\pmb \rho}_{12} \, ,
\end{equation}
 (${\bf R}_{12}={\bf R}_1-{\bf R}_2$, ${\pmb \rho}_{12}=({\bf R}_1+{\bf R}_2)$), where $p_{\parallel}$ does not vanish only for a specific crystallographic nonequivalence of the centers 1 and 2 when there is no inversion center even for collinear Cu\,-\,O\,-\,Cu chain.

The spin-orbital coupling $V_{SO}$  for copper and oxygen ions drives the singlet-triplet mixing which gives rise to a relativistic contribution to electric polarization deduced from an effective spin operator, or an {\it exchange-relativistic-dipole} moment which can be written as follows
$$
{\bf \hat P}_{12}=-\frac{1}{J_{12}}{\bf \Pi}_{12}\left({\bf D}_{12}\cdot[\hat{\bf s}_1\times \hat{\bf s}_2]\right)=
$$
\beq
 -\frac{d_{12}(\theta )}{2l^2J_{12}}(p_{\parallel}{\bf R}_{12}+p_{\perp}{\pmb \rho}_{12})\left([{\bf R}_{12}\times{\pmb \rho}_{12}]    \cdot[\hat{\bf s}_1\times \hat{\bf s}_2]\right)  \, .
\label{rel}
\eeq
In other words, the exchange-relativistic contribution to the dipole moment is a superposition of the two (“longitudinal” and “transversal”)
mutually orthogonal contributions determined only by the superexchange Cu\,-\,O\,-\,Cu geometry, while the “spin-current” factor    does only modulate its value.

The DM type exchange-relativistic dipole moment (\ref{rel}) is believed to be a dominant relativistic contribution to electric polarization in the Cu$_1$\,-\,O\,-\,Cu$_2$  or similar clusters.
It is worth noting that the exchange-dipole moment operator (\ref{non}) and exchange-relativistic-dipole moment operator (\ref{rel}) are obvious counterparts of the Heisenberg symmetric exchange and Dzyaloshinskii-Moriya antisymmetric exchange, respectively. Hence, the Moriya like relation  $|\Pi_{ij}|\sim \Delta g/g |{\bf \Pi}|$  seems to be a reasonable estimation for the resultant relativistic contribution to electric polarization in  M$_1$-O-M$_2$ clusters. At present, it is a difficult and, probably, hopeless task to propose a more reliable and so physically clear estimation.
 Taking into account the typical value of $\Delta g/g \sim 0.1$ we can estimate the maximal value of $|\Pi_{ij}|$ as $10^{-3}|e|\AA$($\sim 10^2 \mu C/m^2$) that points to the exchange-relativistic mechanism to be a weak contributor to a giant multiferroicity with ferroelectric polarization of the order of $10^3 \mu C/m^2$ as in TbMnO$_3$,\cite{KimuraHur} though it may be a noticeable contributor in, e.g., Ni$_3$V$_2$O$_8$\,\cite{Lawes}.

Concluding the section, let us pay attention to the magnetoelectric effect in orthoferrites RFeO$_3$.
 Their $Pbnm$ structure  is particularly
interesting in terms of non-collinear magnetism, however, the perovskite $Pbnm$ phase is not polar and thus none of the compounds crystallizing within this space group will exhibit a spontaneous electric polarization. It can be shown, from pure symmetry arguments, that neither the spontaneous polarization nor the linear or quadratic magnetoelectric effect is possible if the magnetic
order involves the Fe-site spins only.  At the same time,
 spin-canted structures of rare-earth orthoferrites  can produce ferroelectricity, if
 there exists some R-site orderings enabling, by themselves, the linear magnetoelectric effect. For instance, GdFeO$_3$ and DyFeO$_3$ represent two important examples of such low-temperature multiferroics\,\cite{Cano}. Interestingly, the substitution of some of the dysprosium ions in DyFeO$_3$ for bismuth ions leads to the appearance of a strong quadratic magnetoelectric effect, the nature of which is associated with the anomalously high polarizability of Bi$^{3+}$ ions leading to the formation of extended clouds of local electric polarization near Bi$^{3+}$ ions\,\cite{ME}.

\section{Conclusion}

The DM coupling being simple in form, can result in very different magnetic phenomena: weak ferro- antiferro- and weak transversal  ferrimagnetism in a wide number of magnetic 3$d$ oxides, multiferroism, helimagnetism in \Cs \,, helical and skyrmion structures in MnSi-type crystals etc.
In the paper we performed an  overview of the microscopic theory of the DM
coupling and other related exchange-relativistic effects such as exchange anisotropy, electron-nuclear antisymmetric supertransferred hyperfine  interactions, antisymmetric magnetogyrotropic effects, and antisymmetric magnetoelectric coupling
 in orthoferrites RFeO$_3$  and several typical weak ferromagnets.
Most attention in the paper focused on the comprehensive generalization of the Moriya's theory and derivation of the Dzyaloshinskii vector, its value, orientation, and sense  under different types of the (super)exchange interaction and crystal field.   Microscopically derived expression for the dependence of the Dzyaloshinskii vector on the superexchange geometry made it possible to find all the overt and hidden canting angles, that is weak ferro- and antiferromagnetic modes in orthoferrites RFeO$_3$.
The theoretical predictions have been successfully confirmed by various experimental techniques.
   Being based on the theoretical predictions regarding the sign of the Dzyaloshinskii vector we have predicted and studied in detail a novel magnetic phenomenon, {\it weak ferrimagnetism} in mixed weak ferromagnets such as RFe$_{1-x}$Cr$_x$O$_3$ with competing signs of the Dzyaloshinskii vectors.
   In contrast to the end compositions, weak ferrimagnets possess a complex of unusual magnetic properties, including concentration and temperature compensation points, new spin-reorientation transitions, including the newly discovered transition to the angular phase with the spatial orientation of the antiferromagnetism vector and emergence of the $b$-component of the magnetic moment.
   These new materials have broad prospects for practical application, including thermally assisted magnetic random access memories, thermomagnetic switches and other multifunctional devices.

    The  ligand NMR measurements in weak ferromagnets are shown to be an effective tool to inspect the effects of DM coupling in an external magnetic field and determine the mutual orientation of the vectors of ferro- and antiferromagnetism, and therefore establish the sign of the Dzyaloshinskii vector. We considered a number of exchange-relativistic interactions, which in one way or another have a common nature with the spin-bilinear DM coupling.
    As a result of a detailed analysis of the tensor structure of the exchange-relativistic two-ion anisotropy, the emergence of new nondipole contributions with an unconventional temperature dependence has been established.
    It is shown that an analysis of the field dependencies of the $^{57}$Fe NMR frequencies in orthoferrites indicates the existence of a noticeable antisymmetric supertransferred hyperfine interaction as an electron-nuclear analogue of the Dzyaloshinskii interaction.
    A new unusual analogue of the DM coupling, the antisymmetric  spin-other-orbit interaction   can make a decisive contribution to the circular magneto-optics of weak ferromagnets.
We considered the exchange-relativistic antisymmetric contribution to the spin-dependent electric polarization for 3$d$ magnets and established the dependence of the dipole moment of superexchange-coupled ions on the bonding geometry.

The DM coupling as well as related exchange-relativistic effects are nowadays believed to play furthermore a prominent role in many  strongly correlated materials.

I am grateful to I.~E. Dzyaloshinskii for supporting my work and stimulating discussions. I consider it my duty to note that
 most of the work was done in close collaboration with A.~M. Kadomtseva and her laboratory staff.
 I thank  E.~V. Sinitsyn and I.~G. Bostrem for very fruitful multi-year collaboration,  S.~V. Maleev, A.~K. Zvezdin, B.~Z. Malkin, M.~V. Eremin, A.~A. Mukhin, B.~S. Tsukerblatt, S.-L. Drechsler, R.~E. Walstedt, and V.~E. Dmitrienko for stimulating and encouraging discussions.

This research was funded by Act 211 Government of the Russian Federation, agreement No 02.A03.21.0006 and by the Ministry of Education and Science, project No FEUZ-2020-0054.

\end{document}